\documentclass[useAMS,usenatbib]{mn2e}

\usepackage{graphicx}

\title[Tests of element response functions and spectra]{Tests 
of model predictions for the response of stellar spectra and 
absorption line indices to element abundance variations.}
\author[A. E. Sansom, A. de Castro Milone, A. Vazdekis, P.S\'{a}nchez-Bl\'{a}zquez]
{A. E. Sansom$^{1}$\thanks{E-mail:AESansom@uclan.ac.uk (AES)}
, A. de Castro Milone$^{2}$, A. Vazdekis$^{3}$$^,$$^{4}$, 
P. S\'{a}nchez-Bl\'{a}zquez$^{5}$ \\
$^{1}$Jeremiah Horrocks Institute, School of Computing, Engineering and Physical Sciences, University of Central Lancashire, \\
Preston, PR1 2HE, UK\\
$^{2}$ Divis{\~a}o de Astrof{\'i}sica, Instituto Nacional de Pesquisas Espaciais, Av. dos Astronautas 1758, S{\~a}o Jos{\'e} dos Campos, SP 12227-010, Brazil \\
$^{3}$ Instituto de Astrof{\'i}sica de Canarias, C/ V{\'{\i}}a L\'actea s/n, E-38200 La Laguna, Tenerife, Spain \\ 
$^{4}$ Departamento de Astrof{\'{\i}}sica, 
Universidad de La Laguna, E-38205 La Laguna, Tenerife, Spain \\
$^{5}$ Departamento de F{\'i}sica Te{\'o}rica, Universidad Aut{\'o}noma de Madrid, 28049, Spain \\
}
\begin{document}

\date{...Received ...}

\pagerange{\pageref{firstpage}--\pageref{lastpage}} \pubyear{2002}

\maketitle

\label{firstpage}

\begin{abstract}
To analyse stellar populations in galaxies a widely used method is 
to apply theoretically derived responses of stellar spectra 
and line indices 
to element abundance variations, hereafter referred to as response 
functions. These are applied in a differential way, to base 
models, in order to generate spectra or indices with different 
abundance patterns. In this paper sets of such response functions
for three different stellar evolutionary stages 
are tested with new empirical [Mg/Fe] abundance data for the 
MILES stellar spectral library. Recent theoretical models and 
observations are used to investigate the effects of [Fe/H], 
[Mg/H] and overall [Z/H] on spectra, via ratios of spectra for 
similar stars. Global effects of changes in abundance patterns 
are investigated empirically through direct comparisons of 
similar stars from the MILES library, highlighting the impact of 
abundance effects in the blue part of the spectrum, particularly 
for lower temperature stars. It is found that the relative behaviour 
of iron sensitive line indices are generally well predicted by 
response functions, whereas Balmer line indices are not. Other 
indices tend to show large scatter about the predicted mean relations. 
Implications for element abundance and age studies in stellar populations 
are discussed and ways forward are suggested to improve the match with 
behaviours of spectra and line strength indices observed in real stars.

\end{abstract}


\section{Introduction}

Element abundance patterns in galaxies hold vital clues to the formation 
and evolution of their stellar populations. Stellar sources of chemical 
enrichment contribute different abundance distributions on different 
timescales. This provides a potential clock for understanding how the 
integrated stellar population was built up over time and hence the star 
formation history of a galaxy. The power of this technique relies on our 
understanding of the different element abundance contributions, quality 
of the spectroscopic data and on being able to accurately recover 
representative element abundances in integrated stellar populations 
from the available data.

Supernova explosions contribute the main sources of chemical enrichments 
for future generations of stars. SNII explosions enrich the interstellar 
medium (ISM) in a short timescale ($t\le 10^8$ yrs) with a wide range of 
heavy elements (including $\alpha$ elements, iron-peak and r-process elements). 
SNIa explosions enrich the ISM over a much more extended timescale, with 
mainly iron-peak elements, including prompt ($t\sim 10^8$ yrs) and delayed 
($t > 10^8$ to $\sim 10^{10}$ yrs) enrichment \citep{b23,b12,b13}.
Hence the $\alpha$-element to iron ratio is an important indicator of the 
timescale of star formation. Intermediate mass stars contribute to 
the lighter elements 
on relatively long timescale ($t > 10^8$ to $\sim$few $\times 10^9$ 
yrs). Additional contributions may come from red giant stars and cosmic-ray 
reactions, however these are not important for the elements considered in 
the present analysis. Understanding of the relevant element abundance 
contributions and 
timescales is still uncertain in some cases. However the match of models 
to the abundance pattern observed in our own Galaxy \citep{b27,b30,b9},
within a factor of two for most elements up to the iron peak, gives 
some confidence that these contributions are broadly understood. 

Methods of measuring abundance patterns in stellar populations range from 
colours \citep[e.g.][]{b6}, 
which are known to harbour degeneracies \citep[see e.g.][]{b1001,b32}
to broad and narrow spectral features \citep[e.g.][]{b19,b32,b33,b22,b202}. 
Another, related approach is to use
spectral indices from scaled-solar populations to generate proxies for
abundance ratios \citep[][, their fig.~25]{b301}.
There are efforts also at generating full integrated spectra 
\citep[e.g.][]{b5,b18,b11}
and using full spectral fitting \citep[e.g.][]{b31}
for abundance ratio analysis, however 
those methods and models are still under development. 
Information about metallicities, $\alpha$-element to iron ratio 
and sometimes individual element abundances are recovered from 
spectral features. One widely used method is to apply element 
abundance response functions derived from theoretical stellar spectra,
which quantify the changes in line-strength indices to variations of 
individual chemical elements. These response functions are built using 
theoretical model atmospheres, combined with radiative transfer codes and 
extensive line lists of atomic and molecular features. These are applied 
in a differential way, to base models from theoretical or observed spectra 
with standard abundance patterns, in order to generate spectra or indices 
for different abundance patterns \citep[e.g.][]{b282,b25,b2001,b52}.
This can have the advantage of reducing 
problems associated with absolute line strength predictions 
from theory, which are limited by incomplete line and molecular band 
transition information.

Much of the analysis of galaxy abundance ratios in the literature is based on 
the Lick spectral indices 
\citep[with band definitions originally from][- hereafter WO97]{b32,b33}
and response functions for these from 
theoretical stellar spectra (e.g. \citet{b29,b10} - hereafter K05, 
\citet{b7} - hereafter H02, \citet{b24,b11} - hereafter L09). 
Differential application of theoretical models, to empirical 
star or simple stellar population (SSP) indices, is currently thought to be 
one of the best approaches to exploring stellar populations with different 
abundance patterns \citep[e.g. see discussion in][]{b31}. 

In particular the response functions (R) of K05 are widely applied.
Here are some examples. \citet{b14} 
used R from both K05 and H02
to derive ages, metallicities and alpha-element abundances
in globular clusters.
\citet{b21} 
used R from K05, applied 
differentially to the empirical stellar library of \citet{b8},
in order to generate SSP 
models with different abundance patterns. These SSPs have since been used 
in several studies to measure ages and compositions of star clusters and 
galaxies.
\citet{b26} 
used R from K05 
to derive ages and abundances of six elements to investigate chemical 
patterns in globular clusters.
\citet{b1} 
used R from K05 to derive ages 
and [$\alpha$/Fe] ratios of dwarf and giant early-type galaxies.

Examples of the use of other response functions in the literature
include: \citet{b111} 
who used the $\alpha$-enhancement 
dependencies found in L09,
to study the effects of horizontal branch stars and the initial mass 
function on the integrated light of globular clusters.
\citet{b221} 
used response function from 
Worthey et al. (private communication), based on the work of L09, to study
stellar abundance variations as a function of cold and ionised gas content 
in a sample of field early-type galaxies.

In this paper we test the robustness of some of those studies listed 
above that 
attempt to accurately represent the dependence of spectral line 
strengths on differing abundance patterns in stars. We do this by 
testing the response functions, on which those above studies rely, 
on a star-by-star basis, comparing model predictions to empirical observations
of individual stars. This is likely to be one of the cleanest approaches 
to testing the methods used to measure abundance patterns that are most 
widely used in the literature. It has the drawback that real star abundance 
patterns are likely to be more complex than the theoretical models assume,
however, it will provide a grounding for the methods used to measure 
[$\alpha$/Fe]. 

New empirical data for stars are now available, which these response 
functions can be tested against in order to check their accuracy against
real stars. These data are from the Medium-resolution INT Library of Empirical 
Spectra (MILES) \citet{b20} 
- hereafter SB06, \citet{b2}. 
This spectral library consists of 985 stars covering a wide range 
of parameter space in effective temperature $T_{\rm eff}$, surface gravity 
($g$) and metal abundance (characterised by 
[Fe/H]\footnote{[X/H]=log(n(X)/n(H))$_{star}$ - log(n(X)/n(H))$_{sun}$, 
where n(X)/n(H) 
is the number abundance ratio of element X, such as Fe, relative to 
hydrogen.}). 
For 752 of these stars the [Mg/Fe] ratio has been compiled in a catalogue 
\citep[][- hereafter M11]{b15}. 
This compilation 
is based on standardised results from high spectral resolution studies, 
plus new measurements from the medium resolution MILES stellar library, 
calibrated to a standard scale using high resolution measurements.
In this work we make use of [Mg/Fe] measurements as a proxy for all
[$\alpha$/Fe] abundance ratios as a homogeneous nucleosynthetic class 
and compare differential results from these empirical data with 
corresponding differential predictions from theoretical models.

This paper is set out as follows. An overview of current knowledge of the 
effects of differing abundance patterns in stars on their spectral features, 
published response functions and empirical data used in this paper 
are discussed in Section 2. Then the response functions are applied and 
compared to empirical data in Section 3. Effects on spectra 
due to differing abundance patterns are compared for theoretical and 
empirical spectra of stars in Section 4. A discussion of the results is given in
Section 5 and conclusions are given in Section 6.

\section[]{Effects of Abundance Patterns}

\subsection[]{General Considerations}

The chemical and physical conditions of a stellar photosphere are 
imprinted on its emergent spectrum.
The major parameter that defines the overall shape of a photospheric 
spectrum is the effective temperature. Then the abundance pattern, 
surface convection 
and surface gravity also affect its spectrum. In particular, we are 
interested in how the photospheric element abundance pattern affects 
its emergent spectrum. The overall metallicity [Z/H] can affect the continuum
shape as well as absorption line strengths. Iron is the main element being 
analysed in most spectroscopic studies of stars (especially FGK types) 
to quantify the chemical abundance in a photosphere. This is because of the 
existence of a myriad of FeI and FeII lines in the optical range, that are 
measurable at high resolution and that contribute to spectral line strengths 
or narrow band indices at lower resolution. The effects of other elements 
can sometimes be more isolated to particular spectral features, however, 
to accurately measure these effects it is very important to be clear 
about what is meant by the metallicity of the star ([Z/H] or [Fe/H]). 
This is true both for the observations and for the theoretical models 
used to investigate them.

A simplification assumed in recent years, in order to probe beyond overall
metallicity and to uncover the information available in abundance ratios for 
galaxies, is that all the $\alpha$ elements behave in lock-step. This is a 
reasonable approximation based on the observational evidence for some 
$\alpha$ elements from stars in our Galaxy. However, it is not exactly correct 
\citep[e.g.][]{b10001,b152,b51}.
In addition, when handling the metallicity budget in stars, oxygen and 
carbon are important contributors, whose patterns do not follow the 
$\alpha$ elements, iron-peak elements or global metallicity,
but have their own significant contributions \citep[e.g.][]{b131}.
For this reason it is more directly linked to observations if
models predict behaviours of varying abundance patterns at fixed [Fe/H] 
(i.e. a single important element) rather than at fixed [Z/H], which is more 
open to interpretation. Unfortunately this is not always the case and to 
recover changes at fixed [Fe/H] from these models it is necessary to make 
assumptions about how [Z/H], [Fe/H] and other abundance indicators such 
as [$\alpha$/Fe] 
are related. These uncertainties have been more widely discussed in the 
literature in recent years \citep[e.g.][]{b21}
and are emphasised here to clarify the difficulties in
accurately determining abundance patterns from observed stars or stellar 
populations, given the currently available models.

\subsection[]{Response Functions in the Literature}

Response functions tabulate how much various spectral 
line strengths alter with element abundance changes in the theoretical 
model spectra. 
Application of these response functions allows empirical or theoretical 
line strengths to be modified for particular abundance patterns, notably 
enhanced [$\alpha$/Fe] ratios, compared to that of local solar neighbourhood 
stars. Particular response functions in the literature are:

* \citet{b29} 
(TB95) - Models for 3 stars: a cool dwarf, a turn-off 
and a red giant star on a 5 Gyr isochrone. Response functions showed how the 
Lick indices varied due to a factor of 2 increase in individual elements and 
in overall metallicity (i.e. from [X/H]=0.0 to [X/H]=+0.3).

* \citet{b7} 
(H02) - Similar to TB95, but with updated spectral 
line lists, added H$\gamma$ and H$\delta$ indices and carbon enhancements 
reduced to +0.15 rather than +0.3 as used for other elements varied in their 
study (see Worthey 2004 section 3.3). This latter change was an attempt 
to prevent the C$_2$ swan bands from becoming unrealistically strong in 
carbon-rich stars. Their response functions for 3 stars can be obtained from 
http://astro.wsu.edu/hclee/HTWB02). 

* \citet{b10} 
(K05) - Similar to TB95, but for a wider range of initial 
metallicities and star types, with response functions again tabulated for a 
factor of 2 increase in element abundances from the base models.

* \citet{b24} 
(T07) - Generated response functions for a change of $\alpha$ elements 
from [$\alpha$/H]=0.0 (i.e. solar) to [$\alpha$/H]=+0.4. Individual 
elements are not varied, but $\alpha$ elements are enhanced as a group. They start 
from base stars that cover a wider range of atmospheric parameters than in 
TB95, covering up to 5 values of $T_{\rm eff}$ and 4 values of $\log~g$.
T07 do not give responses for overall changes in metallicity.
These response functions have not yet been widely used subsequently in the literature. 

* \citet{b11} 
(L09)- Expanded the work of H02 and generated 
response functions for SSPs using many ($\sim35$) theoretical star 
spectra at solar metallicity times 10 individual element enhancements 
(at fixed overall metallicity). 
Their theoretical spectra are binned to 0.5\AA\ per flux point 
(however their response functions are not very sensitive to spectral 
resolution). Plots of some comparisons with K05 for individual 
theoretical stars are given at http://astro.wsu.edu/hclee/NSSPM$\_$Lick.html; 
these show similar, but not identical, responses in general between K05 
and their evaluations.

Some of the above response functions varied the amounts of individual 
elements present in the atmospheres, however, they did not always track
changes in opacity self consistently. For example, K05 tracked opacity 
changes for overall metallicity changes ([Z/H]), but treated individual 
elements like trace elements, whereas the theoretical spectra of L09 
were consistently calculated for each abundance pattern.
More recent theoretical spectra are available that take into account 
non-solar abundance patterns plus a more self-consistent 
approach \citep[e.g.][]{b501,b151}. 
In particular the theoretical stellar spectra of \citet{b501} 
are compared to observational spectra in 
Section 4 of this paper. Response functions for Lick indices are not 
generally available for these recent theoretical stellar libraries. 

Table~1
shows basic characteristics, assumptions and tools used in the 
generation of published element response functions for stars. This shows the
range of different models and assumptions used in generating these response
functions.

\begin{table*}
 \centering
 \begin{minipage}{175mm}
\begin{center}
  \caption{Table showing basic assumptions and tools used in the 
generation of published element response functions for stars. 
Elements listed are those tabulated in the response functions. See the 
Author references in column 1 for details of other references and names
given in this table.}
  \begin{tabular}{llllcc}
  \hline
   Author & Stellar    & Spectral   & Other    &  $\alpha$ elements & Other \\
          & Atmosphere & Synthesis  & Comments &                    & elements \\
          & Code       & Code       &          &                    &  \\
  \hline
 H02     & MARCS       & SSG (Bell \& Gustafsson 1989) & Updated TB95 & O,Mg,Si,Ca,Ti & C,N,Na,Cr,Fe \\
 K05     & MAFAGS      & LINFOR                      & Excludes TiO & O,Mg,Si,Ca,Ti & C,N,Na,Cr,Fe \\
 T07     & ATLAS9      & Munari et al 2005           & Combined $\alpha$ & $\alpha$-enhancement & [Z/Z$_{\sun}$] \\ 
 L09     & Plez        & FANTOM (Coelho et al. 2005) & Coolest stars & O,Ne,Mg,Si,S,Ca,Ti & C,N,Fe \\       
     ``  & ATLAS       & FANTOM (Coelho et al. 2005) & Cool stars    & & \\
     ``  & MARCS       & SSG (Bell et al. 1994)      & Medium $T_{\rm eff}$   & & \\
     ``  & ATLAS       & SYNTHE (Kurucz 1970)        & Hot stars     & & \\
\hline
\end{tabular}
\end{center}
\end{minipage}
\label{model_assumptions}
\end{table*}

\subsection[]{Observations: MILES Lick Line Strength Indices}

For the 752 stars for which [Mg/Fe] could be obtained in \citet{b15} 
we measured line-strength indices in the Lick/IDS system 
(with the definitions of \citet{b281} and WO97)
in the latest version of the MILES stellar spectra \citep{b53}. 
Errors were 
estimated from uncertainties caused by photon noise and
wavelength calibration (errors in the flux calibration were not taken 
into account, but the relative flux calibration in the MILES stars 
have been proved to be very accurate). The line-strength indices 
were transformed to the Lick system taking into account differences 
in spectral resolution between the Lick/IDS system and MILES stars 
following the prescriptions in WO97 (their table~8). The final 
resolution at which each index was measured is given in Table~2.
No further offsets were applied to the measured indices, since both 
the theoretical response functions and MILES observations were 
not converted to the Lick/IDS flux system (see K05 section 2.4).
Average errors and units for each index are given in the last two columns 
in Table~\ref{lick_res}. 
Appendix A lists all the parameters and Lick indices for MILES stars 
used in Figs.~1 and 2 of this paper.

\begin{table}
\caption{Lick resolution and MILES average errors. First column gives 
the index name; 
second column gives the final spectral resolution (FWHM) at which each 
index was measured; third column gives the average Lick index errors for 
the MILES stellar database and their units are given in column 4.}
\begin{tabular}{lrll}
\hline
Index       & Resolution & Index       &  Index \\
            & (\AA)      & Ave. Error  &  units \\
H$\delta_A$ & 10.9 & 0.1895 & \AA \\      
H$\delta_F$ & 10.9 & 0.1278 & \AA \\      
CN$_1$      & 10.6 & 0.0050 & mag. \\       
CN$_2$      & 10.6 & 0.0061 & mag. \\      
Ca4227      & 10.1 & 0.0850 & \AA \\      
G4300       & 9.8  & 0.1427 & \AA \\      
H$\gamma_A$ & 9.5  & 0.1533 & \AA \\      
H$\gamma_F$ & 9.5  & 0.0912 & \AA \\       
Fe4383      & 9.2  & 0.1921 & \AA \\       
Ca4455      & 9.1  & 0.0970 & \AA \\
Fe4531      & 9.0  & 0.1377 & \AA \\
C$_2$4668   & 8.8  & 0.1942 & \AA \\
H$\beta$    & 8.4  & 0.0740 & \AA \\
Fe5015      & 8.4  & 0.1528 & \AA \\
Mg$_1$      & 8.4  & 0.0016 & mag. \\
Mg$_2$      & 8.4  & 0.0018 & mag. \\
Mgb         & 8.4  & 0.0653 & \AA \\
Fe5270      & 8.4  & 0.0692 & \AA \\
Fe5335      & 8.4  & 0.0698 & \AA \\
Fe5406      & 8.4  & 0.0505 & \AA \\
Fe5709      & 9.2  & 0.0890 & \AA \\
Fe5782      & 9.2  & 0.0849 & \AA \\
NaD         & 9.5  & 0.1103 & \AA \\
TiO$_1$     & 9.7  & 0.0026 & mag. \\
TiO$_2$     & 9.7  & 0.0023 & mag. \\
\hline
\end{tabular}
\label{lick_res}
\end{table}

\begin{table*}
 \centering
 \begin{minipage}{175mm}
\begin{center}
\caption{Parameters for base stars that are used for normalisations in 
the response function tests. The empirical parameters listed are for MILES base 
stars that match the base stars modelled by K05, within observational errors. 
These three stars are also modelled by H02. Maximum offsets assumed 
for this match are: $\Delta$T=$\pm$100K, $\Delta\log~g$=$\pm$0.2, 
$\Delta$[Fe/H]=$\pm$0.1 and $\Delta$[Mg/Fe]=$\pm$0.06. The final column lists 
reference sources for the model or observation and also indicates the type of data 
available for [Mg/Fe] determinations for each base star (see M11 for details).
 }
\begin{tabular}{llccrrl}
\hline
Star  & Model or    & $T_{\rm eff}$ & $\log~g$ & $[Fe/H]$ & $[Mg/Fe]$ & Source \\
Type  & Star Name   &   (K)   &        &        &         &        \\
\hline
CD    & K05 table 12    & 4575    & 4.60   &  0.00  &  0.00   & K05 \& H02 \\ 
CD    & HD032147    & 4658    & 4.47   & +0.02  & -0.06   & M11 (HR) \\ 
 &&&&&& \\     
TO    & K05 table 13    & 6200    & 4.10   &  0.00  &  0.00   & K05 \& H02  \\ 
TO    & HD016673    & 6253    & 4.28   & +0.05  & +0.05   & M11 (HR) \\ 
 &&&&&& \\
CG    & K05 table 14    & 4255    & 1.90   &  0.00  &  0.00   & K05 \& H02  \\      
CG    & HD154733    & 4200    & 2.09   &  0.00  & -0.03   & M11 (mrBothMg) \\ 
\hline
\end{tabular}
\end{center}
\end{minipage}
\label{base_stars}
\end{table*}

\section[]{Testing Response Functions}

The original study of TB95 opened the way to differential techniques for
tracking abundance ratios. Their work was followed by the more comprehensive 
study of K05, who included the same theoretical base, solar metallicity stars 
as TB95, but also explored response functions from lower and higher metallicity 
base stars. K05 showed that their work agreed well with the results of TB95 and 
also added results for H$\gamma$ and H$\delta$ indices. Response functions from 
both these studies have been widely applied in the literature, however we do not
test the TB95 response functions here since those of K05 and H02 can be 
considered as expansions of that earlier work.
Theoretical spectra from H02 were revised and expanded on by L09 to form SSPs 
and their theoretical spectra and SSP response functions are used in the 
literature, mainly by that group.
In this section we test the publicly available star response functions 
of K05 and H02, which are the ones most widely applied in the literature 
that we can test. We also test the star response functions used by L09, 
from information provided by G. Worthey (private communication).

It is important to note that the published tables of response functions 
tested in this paper tabulate responses of Lick spectral indices to changes 
in abundance of individual elements, treating individual elements as trace 
abundances and assuming that the opacity distribution in the atmosphere is not 
significantly altered by changing the abundance of one element. However, 
the response function tables (of TB95, K05 and H02) also tabulate changes 
in indices due to changes in overall metallicity [Z/H] and those values 
do take into account changes in the structure in the stellar atmospheres 
due to opacity changes. Since iron is a very important opacity source in 
stars, and is also most generally the element abundance measured in 
libraries of stars, we make use of these more self-consistent changes 
due to [Z/H] to go from theoretical base star indices to stars with 
different [Z/H]=[Fe/H] values, then we further adjust these index 
changes to account for non-solar abundance ratios relative to iron, 
treating the other elements (mostly $\alpha$ elements) as trace element 
changes. In this way we aim to make best use of the physics that went 
into the models. For further discussion about the different order in 
which the response function tables might be applied, see \citet{b191}.

\subsection{K05 Response Functions}

The response functions of K05 were generated from theoretical
spectra blurred to the resolution of the Lick/IDS system, according to the 
resolution variations with wavelength measured by WO97 (see K05 section 2.4). 
Other corrections to the Lick/IDS standard system were not applied, which is 
the same approach as for the empirical observations that we are using here. 
Any remaining differences due to continuum normalisation will be second 
order effects, mainly affecting the broader band indices.  
The differential approach used in applying response functions will
reduce the need for corrections due to differences in flux calibrations.
However, we note that the broader spectral features are the ones most likely
to show residual affects due to any remaining flux offsets. 
These are the CN, Mg and TiO bands.  

\subsubsection{Solar abundance pattern base models}

Although K05 presented response functions for base star models with 
different metallicities and some different abundance patterns, only 
those for base stars with solar abundances and solar abundance ratios
can be tested here. This is because there are no suitable observed 
stars in the MILES library to match the specific base star models 
tabulated in K05 with non-solar abundances. That is, only tables 12-14 
(5Gyr models) and tables 15 \& 17 (1Gyr models) in K05 have sufficient 
matching stars in MILES to be able to test them.

The theoretical model values are first derived from K05, using their 
tables 12, 13 and 14, which give element dependent response functions 
for a cool dwarf (CD), a turn-off (TO) and a cool giant (CG) star 
respectively, from a 5 Gyr populations, each of which starts with 
solar metallicity and abundance ratios. Base star parameters for these 
three models are shown in Table~3
together with matched observations used to make the normalised comparisons.
We apply the response functions twice to the base models; once to generate 
a set of theoretical indices for the correct [Fe/H] for the star being modelled,
by generating indices for a star with that overall [Z/H] (initially 
with solar abundance ratios), then again to modify 
those theoretical indices to the correct [$\alpha$/Fe] of each of the observed 
stars being modelled. We enhance the $\alpha$ elements listed in Table~1,
together with Na (see K05, Section 2.1), whilst C, N, Cr and Fe remain 
un-enhanced. We use this two step process since there are 
insufficient observed stars of the specific $T_{\rm eff}$, $\log~g$, [Fe/H] 
combinations modelled, which could be used to isolate only [$\alpha$/Fe] 
enhancement effects.

\begin{figure*}
 \centering
 \begin{minipage}{140mm}
\begin{center}
 \includegraphics[width=91mm, angle=0]{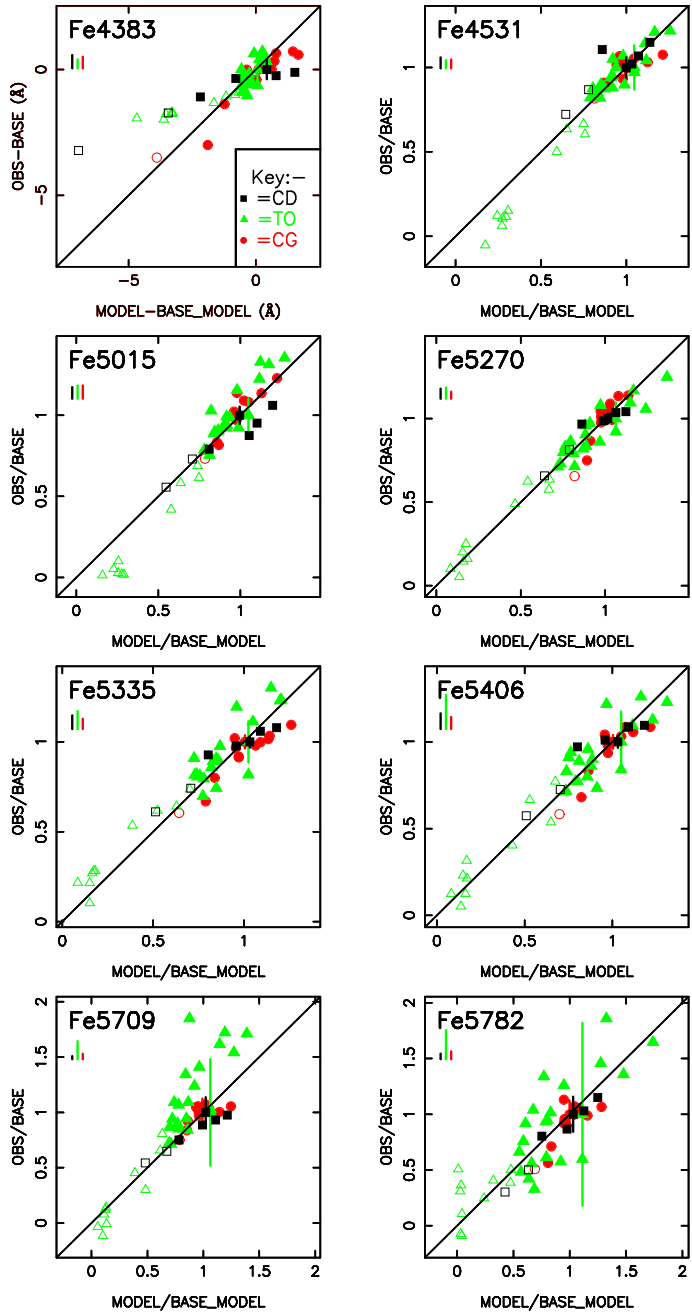} 
 \caption{\hskip -5pt{\bf (a).} Testing the response functions of K05. 
Comparison of normalised empirical versus normalised theoretical line 
strengths for standard Lick indices sensitive to Fe lines in the stellar 
photospheres. (Note that Fe4531 and Fe5015 are more sensitive to overall 
metallicity [Z/H] - see K05).
The empirical observations are for stars in the MILES spectral library, 
with known [Fe/H] and [$\alpha$/Fe] (from [Mg/Fe] measurements in M11).
Three star types are shown: cool dwarfs (CD, black squares), 
turn-off stars (TO, green triangles) and cool giants (CG, red circles). 
The observed stars shown are chosen to all have the same $T_{\rm eff}$ and 
$\log~g$ as the base theoretical star within observational errors, 
for these three categories. The observed base stars used are 
HD032147 (CD), HD016673 (TO), HD154733 (CG), which match the 
atmospheric parameters tabulated for these three star types with solar 
abundance pattern in K05 (their tables 12, 13 and 14 respectively). 
Both empirical (vertical) and theoretical (horizontal) axes show either 
{\it differences} (for molecular bands and for lines indices that go 
negative or close to zero) or {\it ratios} (for indices that stay positive 
for all stars).
Average observational errors (2$\sigma$) are attached to the base star points 
for each index. Systematic errors, estimated from observational uncertainties 
in base star atmospheric parameters, are indicated by the three vertical lines 
under the index name on each plot. These are colour coded for each star type 
modelled. For differences, these systematic error bars represent the maximum 
vertical offset expected due to combined line-strength uncertainties (added 
in quadrature), from uncertainties in $T_{\rm eff}$, $\log~g$ and [Fe/H]; 
while for ratios, these error bars indicate the slope uncertainty 
at one on the vertical axis due to these combined uncertainties.
The straight line shows the one-to-one relation in each case. 
Open symbols show stars with [Fe/H]$<$-0.4, indicating stars with 
much lower metallicities than the base stars.}
\end{center}
\end{minipage}
\end{figure*}

The observed stars chosen for the comparison are selected to be those that 
have the same atmospheric parameters of effective temperature and 
surface gravity as the tabulated theoretical model stars of K05, 
within the observational errors on these parameters. 
For the three base parameters, these errors are $\Delta T$=$\pm 100$ K, 
$\Delta\log~g$=$\pm 0.2$ and $\Delta$[Fe/H]=$\pm 0.1$ dex.
For [Mg/Fe] we choose stars within $\Delta$[Mg/Fe]=$\pm 0.06$ dex, 
since this is the main parameter that we are testing.
Only specific base star models have response functions tabulated in K05,
therefore that determines our choice of stars that we can test.
The observed $T_{\rm eff}$ and $\log~g$ values are those given in the MILES 
spectroscopic database 
\citep{b2},
the observed [Fe/H] and [Mg/Fe] values are those given in M11
and the observed line strengths used 
are measured from MILES spectra convolved to the same spectral resolution 
as in K05 (as described in section 2.3 and tabulated in Appendix A).
Ratios (or differences) are then formed for both the 
observations (S) and corresponding theoretical model (M) indices.
For a perfect match between observations and response function predictions 
the ratio M/S would equal 1 (or differences would equal zero).

\begin{equation}
   \rmn{Ratio}=M/S
\end{equation}

where: 
 $$M=\frac{M_*([Fe/H],[\alpha/Fe])}{M_*(0,0)}$$  is the theoretical model ratio, and 
 $$S=\frac{S_*([Fe/H],[\alpha/Fe])}{S_*(0,0)}$$  is the observed star ratio, 
where the denominators in $M$ and $S$ are the base star values.

Equations used to correct for different abundance patterns using the tabulated 
responses are from \citet{b25} 
(their equation 7) and 
K05 (their equation 3), for index and flux corrections 
respectively.  The equations used are described below.

Fractional changes in indices ($\Delta I/I_0$), due to the combined 
effects of tabulated response functions $R(i)$ for elements $i$=1 to $n$, 
from \citet{b25},
their equation 7 are:-
\begin{equation}
   \frac{(I_{new}-I_0)}{I_0}=\frac{\Delta I}{I_0}=\Pi_{i=1}^n(exp[R_{0.3}(i)])^{(\Delta[x_i]/0.3)} - 1
\end{equation}
where $R_{0.3}(i)$ are the tabulated fractional index changes 
for a factor of 2 increase in abundance of element $i$, and $\Delta[x_i]$ is the 
change in logarithmic abundance of element $i$ (i.e. $\Delta[x_i]=+0.3$ for a 
factor of 2 increase in abundance of element $i$) .

Fractional changes in line fluxes ($\Delta F_l/F_{l0}$), due to the combined 
effects of index changes, 
from K05, their equation 3 are:-
\begin{equation}
   \frac{(F_{lnew}-F_{l0})}{F_{l0}}=\frac{\Delta F_l}{F_{l0}}=\Pi_{i=1}^n(exp[\frac{\delta F_l}{F_{l0}})^{(\Delta[x_i]/0.3)} - 1
\end{equation}
where $\delta F_l/F_{l0}$ is the flux change for a factor of 2 increase 
in abundances of element i.

Index and flux are linked via the equation:-
\begin{equation}
   I=W(1-\frac{F_l}{F_c})
\end{equation}
where $F_c$ is the continuum flux and $W$ is the bandwidth for index $I$.
which leads to:
\begin{equation}
   \frac{\delta F_l}{F_{l0}}=\frac{\delta I}{(I_0-W)}
\end{equation}
as in equation 2 of K05, where $R=\delta I/I_0$ can be obtained from the 
tabulated response functions for specific elements and indices. Equation 5 
can be used in general to convert from flux changes to index changes.

The corrections to indices are applied for those indices that 
behave as expected for weak lines (tending to zero strength for the weakest 
measurements), whereas corrections to fluxes are applied 
when the defined indices can take positive or negative values. This is to 
ensure that the property being corrected for element abundance pattern remains 
positive. After corrections are applied, fluxes are converted back to indices 
in order to make the comparisons with observations.

Fig.~1(a to e) shows comparison plots for the response functions of K05. 
The stars plotted in these figures have a wide range in abundance pattern,
covering $-2.86<$[Fe/H]$<+0.41$ and $-0.10<$[$\alpha$/Fe]$<+0.53$. 
Stars with [Fe/H]$<$-0.4 are plotted as open symbols to highlight 
extrapolations to low metallicity, away from the base star model 
of [Fe/H]=0.0. 

To assess the significance of differences between observations and models, 
reduced chi-squared values were computed. Some systematic offsets from a 
one-to-one line in the comparison plots are expected due to slight mismatches 
between observed and theoretical base star parameters (see Table~3)
This is unavoidable, since we have a finite number of observed stars 
and a finite number of base models for which theoretical response functions 
are available, and the two do not match perfectly. From the few suitable base 
stars available, it is found that these systematic 
offsets are generally small (typically less than twice the average errors 
on line strengths). They are larger for molecular band features, causing
systematic shifts of up to $\pm 0.03$ magnitudes away from the one-to-one 
lines in the comparison plots. To estimate the size of stystematic offsets 
expected due to uncertainties in atmospheric parameters of size
$\Delta$T=$\pm$100K, $\Delta\log~g$=$\pm$0.2 and $\Delta$[Fe/H]=$\pm$0.1, we
used the MILES on-line interpolator\footnote{based on real stars, 
http://miles.iac.es/pages/webtools/star-by-parameters.php} 
to generate Lick indices for base stars, varying the parameters by 
these amounts. The average offsets in one direction are shown in 
Figure 1, below the index name in each plot. These are shown for each of 
the three star types tested and represent a maximum typical systematic 
offset expected due to uncertainties in line strength, added in quadrature, 
due to uncertainties in all three atmospheric parameters.
For comparisons shown as differences, any inaccuracy in base star parameters
will appears as a systematic offset above or below the one-to-one line in the 
comparison plots. For comparisons shown as ratios, any inaccuracy in base 
star parameter will appear as a systematic fractional difference.

In order to generate error normalisations for evaluating 
chi-squared, average ($2\sigma$) errors from MILES Lick indices were added 
in quadrature with mean offsets from the one-to-one line, for each index 
and each star type. This will account for offsets due to parameter inaccuracies 
in the base star, but not in the other stars, since the effect of such 
inaccuracies on Lick indices will be random rather than systematic. 
The reduced chi-squared ($\chi_{\nu}^2$) was found by 
dividing by the number of stars in each case, since no parameters were 
being fitted as the comparison is with the one-to-one line prediction.

The results of the comparisons are described in the next section and 
the derived $\chi_{\nu}^2$ values are given in Table~4

\subsection{K05 Results}

Fig.~1(a) shows the results for Lick indices mainly sensitive to Fe or 
overall metallicity. These indices show the expected behaviour for observed 
line-strength changes compared to theoretical ones. There are good one-to-one 
relations for the differential changes plotted between observations and those 
derived from theoretical response functions, given the observational errors. 
The agreement is confirmed by the reduced chi-squared values for these indices,
which are typically $\chi_{\nu}^2<3$ (see Table~4). 
Note that conservative 
2-sigma error bars are plotted for the random Lick measurement errors, 
therefore they look larger than the typical data scatter for weak indices 
such as Fe5782, where this Lick measurement error dominates the scatter.
This agreement is not so surprising for features dependent mainly on Fe or 
overall metallicity, since these dominate spectral changes due to composition 
changes. Both systematic and random errors are generally larger for 
TO stars, since metal sensitive line strength are generally weak 
(and particularly sensitive to temperature uncertainties) in these warmer 
stars. Other systematic errors are relatively small, consistent with the 
good one-to-one relations seen in this figure.

\addtocounter{figure}{-1}

\begin{figure}
 \centering
 \includegraphics[width=84mm, angle=0]{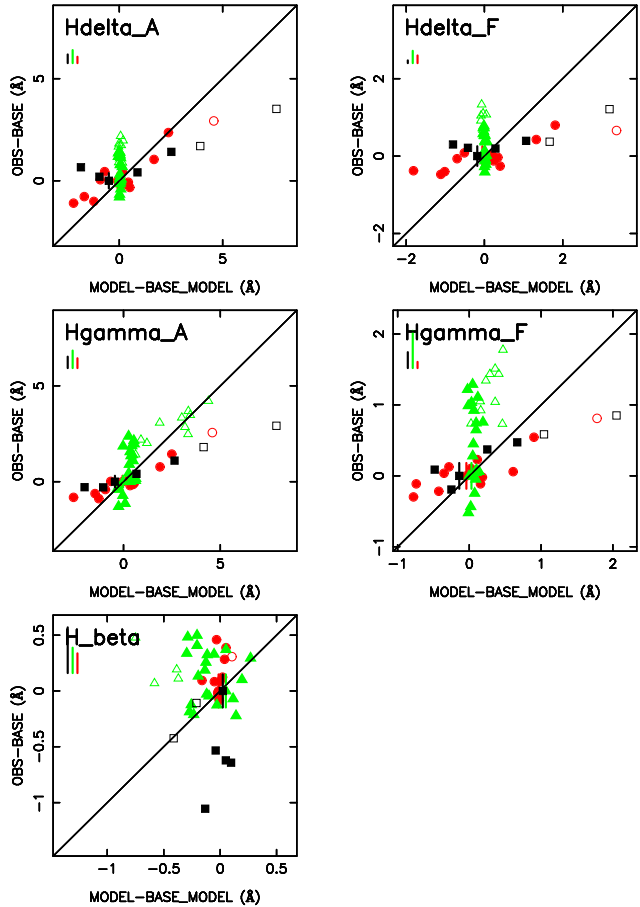} 
 \caption{\hskip -5pt{\bf (b).} Testing the response functions of K05. 
Comparison of normalised empirical versus normalised theoretical line 
strengths for standard Lick indices sensitive to H Balmer lines in the stellar 
photospheres. Symbols as in Fig.~1(a), with cool dwarfs (CD, black squares), 
turn-off stars (TO, green triangles) and cool giants (CG, red circles).}
\end{figure}

\addtocounter{figure}{-1}

\begin{figure}
 \centering
 \includegraphics[width=84mm, angle=0]{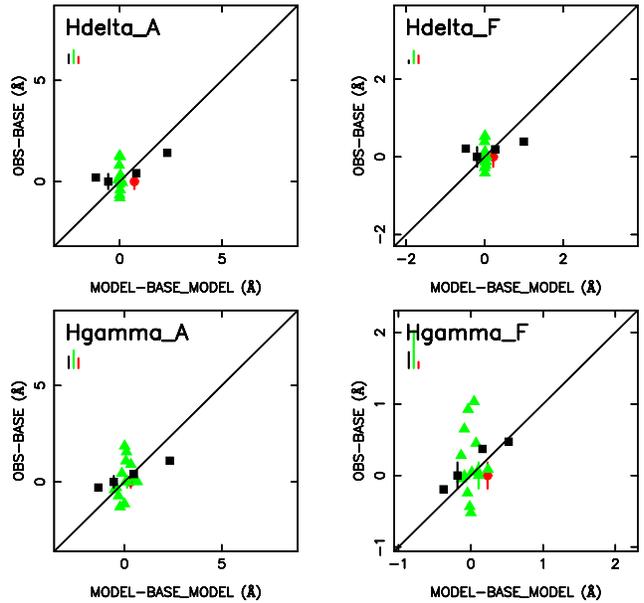} 
 \caption{\hskip -5pt{\bf (c).} The same as Fig.~1(b) top 4 panels,
but for a restricted set of tested stars from MILES, which also have individual
C,N and O abundance measurements applied.}
\end{figure}

\addtocounter{figure}{-1}

\begin{figure}
 \centering
 \includegraphics[width=84mm, angle=0]{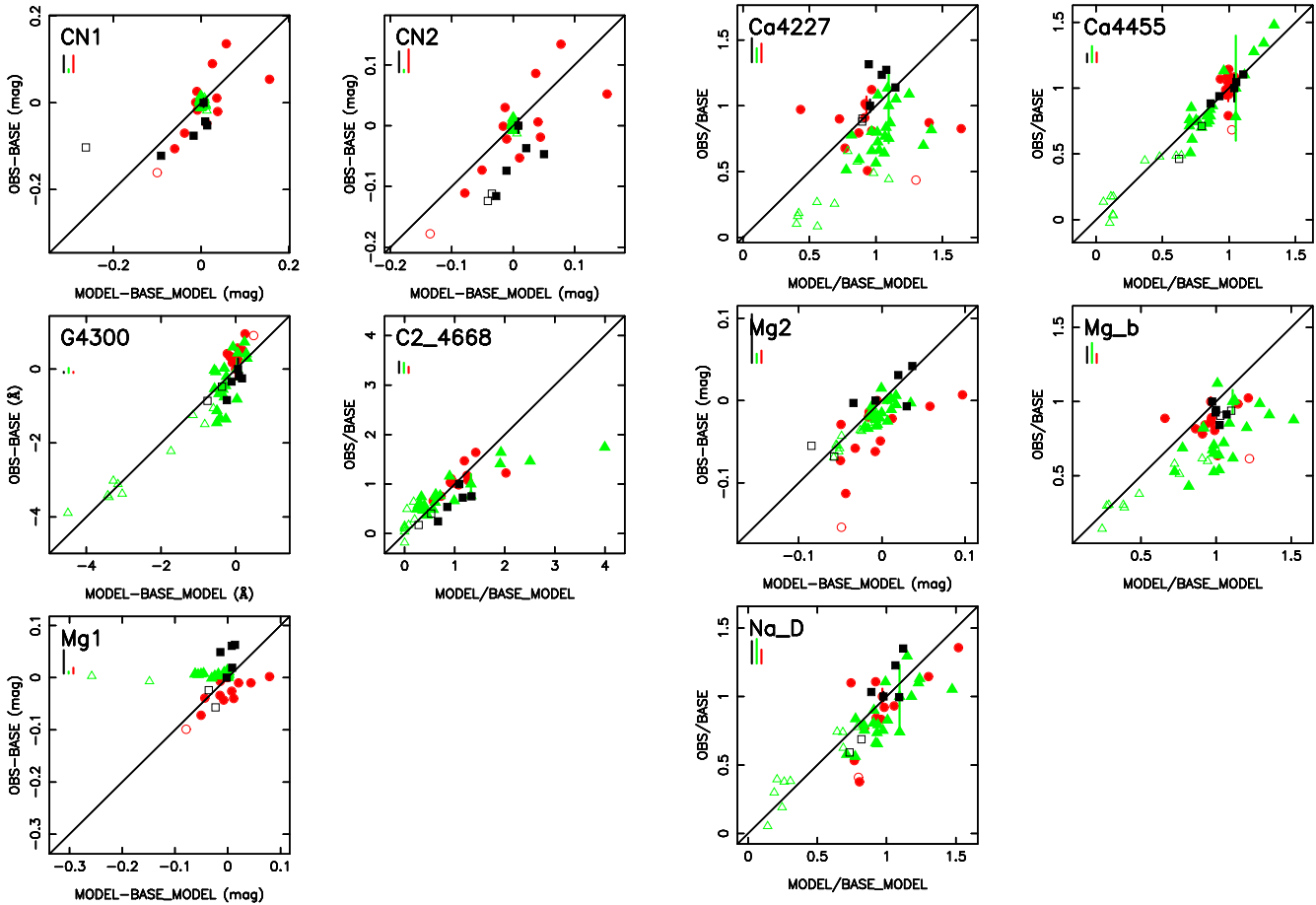} 
 \caption{\hskip -5pt{\bf (d).} Testing the response functions of K05. 
Comparison of normalised empirical versus normalised theoretical line 
strengths for standard Lick indices sensitive to light metal 
(C,N and O) elements 
in the stellar photospheres. Symbols as in Fig.~1(a).}
\end{figure}

\addtocounter{figure}{-1}

\begin{figure}
 \centering
 \includegraphics[width=84mm, angle=0]{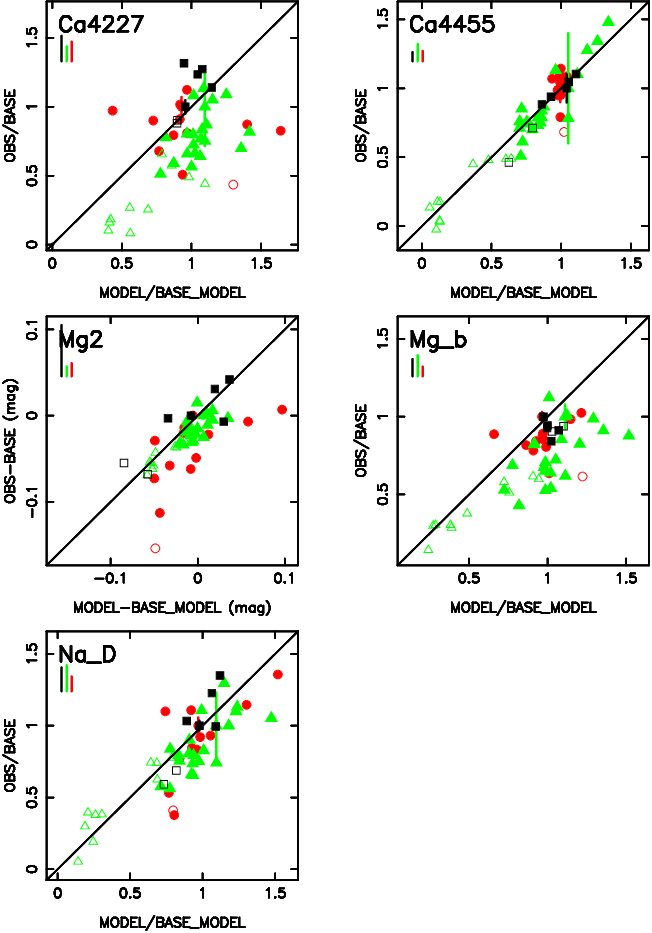} 
 \caption{\hskip -5pt{\bf (e).} Testing the response functions of K05. 
Comparison of normalised empirical versus normalised theoretical line 
strengths for standard Lick indices sensitive to other elements 
in the stellar photospheres (e.g. Ca4227 is most sensitive to Ca;
Ca4455 is weakly sensitive to a variety of elements;
Mg$_2$ and Mgb are most sensitive to Mg;
NaD is most sensitive to Na). Symbols as in Fig.~1(a).}
\end{figure}

Fig.~1(b) shows results for H-Balmer Lick indices. For H$\gamma$ and H$\delta$ 
the K05 response functions do not mimic well what is happening in real 
stars as a function of changing abundance patterns ([Fe/H] and [Mg/Fe]). 
In the CD and CG stars
the theoretical response functions predict larger changes than are observed 
in the empirical star data. For the TO stars, the reverse is true, and
the K05 response functions predict negligible variations in these indices 
as a function of changing abundance patterns (as highlighted, for example 
in the vertical column of green triangles in the H$\delta_A$ plot of 
Fig.~1(b)). Observed variations of H$\gamma$ and H$\delta$ indices in 
these warmer stars are larger than the theoretical response functions 
predict. Some of the vertical scatter in these plots will result from
inaccuracies in parameter measurements from star-to-star. Inaccuracies in 
base star parameters are not the cause of the systematic differences between 
observations and predictions, since that would cause systematic offsets 
rather than changes in slope around the mean, as observed in Fig.~1(b).
This is confirmed by the relatively small systematic error bars for the 
cool stars, shown below the index labels on each plot.
The differences between observations and predictions are highlighted 
by the large $\chi_{\nu}^2$ values for CG and CD stars, for these higher 
order Balmer indices (see Table~4). 

H$\gamma$ and H$\delta$ indices are known to be affected by 
CN bands within the definition of these indices. H$\gamma_A$ might be 
sensitive to CH (i.e. G band) affecting its blue pseudocontinuum, whilst
CN at 4150\AA\ might affect the red pseudocontinuum. Therefore the difference, 
in principle, could be due to differences in C and N abundances, with
carbon effects being particularly important in the response functions.
Linear fits to the cool stars data ([Fe/H]$\ge$-0.4) for H$\gamma$ and 
H$\delta$ features in Fig.~1(b), give offsets that imply carbon abundance 
changes much larger than the maximum observed deviations in [C/Fe], 
which are $<$$\pm$0.4 dex (e.g. Luck \& Heiter 2006, 2007). For example, 
for CG stars in H$\gamma_A$ a shift of +1.76\AA\ would bring the lower 
point onto the 1:1 line, but this requires a change in [C/Fe] of 
1.28 dex. Therefore, the slopes for cool stars in Fig.~1(b) cannot 
be reconciled with the 1:1 line by appealing to systematic changes 
in carbon abundance alone. Other aspects need to be considered.
We searched the literature for individual measurements of 
C, N or O abundances, relative to Fe, for the stars tested in 
Fig.~1(b) and found only a few. Fig.~1(c) shows the results of 
applying these individual abundance measurements. For the cool stars 
there were only measurements of C and O for 4 CD stars \citep{b1111}
and C, N and O for 1 CG star \citep{b1112}. 
Abundances of C, N and O were available for 14 of the TO stars
\citep{b231},
which are also plotted. In Fig.~1(c) the response functions 
from K05 are applied as for Fig.~1(b) except that 
the columns for responses to individual C, N and O abundances 
are applied, where available, rather than their assumed links 
to [$\alpha$/Fe] and [Fe/H] made previously, in Section 3.1.1.
The systematic slope difference from the 1:1 line, for 5 
cool stars, is still evident in Fig.~1(c), and more measurements 
on individual C, N and O abundances are needed in future for 
further tests of the significance of the effects of these 
individual elements on the higher order Balmer features.

We find in Appendix B that fine 
tuning the abundance ratios, to take into account mean trends of several 
elements, makes no significant difference to the mismatch for 
H$\gamma$ and H$\delta$ indices, using the K05 response functions.
This points to overall metallicity response as the cause of the 
non-unity slopes for cool stars in these four indices (Fig.~1(b)).
For H$\beta$ the scatter is larger than expected from observational 
uncertainties in line strengths, since this index is thought 
not to be very sensitive to chemistry \citep[e.g. L09,][based on the 
synthetic stellar library of Coelho et al. 2005]{b201};
however, note that a different conclusion for H$\beta$ was reached 
by \citet{b24,b201}, 
based on the synthetic stellar library of \citet{b151}.
The offsets seen in H$\beta$ can be explained by the systematic error 
bars plotted, which are particularly large for the CD stars in this index.

Fig.1(d) shows Lick indices that are particularly sensitive to the light 
metals, C, N and O. These behave qualitatively as expected from the response 
functions, but with larger scatter than expected from the line-strength 
measurement errors in most cases. The $\chi_{\nu}^2$ values reflect 
this (see Table~4).
The lack of variation in the CN indices in TO stars, predicted from the 
theoretical response functions, agrees with the observations.
For these CN indices in cool stars, errors due to atmospheric parameter 
uncertainties contribute to the scatter and offsets.
Differences in CN band strengths between stars is also likely to 
contribute to this scatter. The Mg$_1$ feature, which is most sensitive 
to carbon, varies far more in the theoretical predictions than in the 
observations for the TO stars.
Mg$_1$ in these warm TO stars is very weak compared to its values in the 
CD and CG stars and is observed to vary very little from star-to-star.
Therefore its theoretical response function is uncertain.
Also, predicted ratios for C$_2$4668 extend to higher values for 
some TO stars than in the observations, which do not go above twice 
the base star line-strengths in these stars. 
For C$_2$4668, in TO stars, the response function predictions start 
to deviate significantly for applications to higher metallicities 
([Fe/H]$>$+0.2), where the theoretical predictions have larger 
line-strengths than the stellar observations, by up to a factor of 2, 
as seen in the extreme right TO star in Fig.1(d) for the C$_2$4668 
index. However, we note that in luminous elliptical galaxies this index 
can take higher values (e.g. \citet{b34}, their fig.~14).
Therefore, except for the weak Mg$_1$ feature in TO stars and for C$_2$4668 
in TO stars, the response 
functions are not systematically biased in their predictions for these 
C,N,O sensitive Lick indices (CN$_1$, CN$_2$, G4300, C$_2$4668 and Mg$_1$).
Future high resolution spectroscopic observations are needed to test the 
effects of C and N abundance variations on a star-by-star basis.

Fig.~1(e) shows Lick indices sensitive to other elements, including 
sodium and various $\alpha$ elements (Mg, Ca). Again the response function
predictions are approximately followed by the observation, but with large 
scatter, and some offsets between star types. These systematic offsets 
shift vertically somewhat when different base stars are used, illustrating 
the sensitivity of these features to exact photospheric parameters, 
even within their observational uncertainties. 
The systematic error bars due to atmospheric parameter uncertainties
are relatively large for most of these indices, as seen in this figure.
The calcium sensitive index Ca4227 shows large scatter about the 
one-to-one line, which may also reflect CN contamination effects
\citep{b21}
and/or calcium variations that are not fully in step with magnesium 
variations in the observed stars, since we are using [Mg/Fe] as a proxy 
for all $\alpha$ elements (O, Ne, Mg, Si, S, Ar, Ca, Ti). In fact there 
is evidence that calcium does not follow exactly in step with magnesium 
in a range of environments (e.g. in our Galaxy 0.0$<$[Ca/Fe]$<$[Mg/Fe] 
for metal-poor stars \citet{b51} 
and calcium is even lower [Ca/Fe]$\le$0.0 for open clusters, 
\citet{b171}. 
Also in luminous elliptical galaxies 
calcium appears to follow iron rather than other $\alpha$ elements such 
as magnesium \citep[e.g.][and references therein]{b302,b222}. 
In future it may be possible to compile 
calcium abundances onto a single scale, for a substantial number of 
MILES stars. Such a [Ca/Fe] catalogue would allow us to test whether 
differences in [Ca/Mg] are contributing to the scatter for some of 
the indices. This is particularly important for Ca4227, which 
currently shows poor agreement between response function predictions 
and real stars. Ca4455 is weakly 
sensitive to a number of elements and the observations follow the 
response function predictions well for this Lick index 
(see Table~4),
with CG stars showing only small variations in both the observations and
predictions. 

The Mg$_2$ and Mgb indices, which are sensitive to magnesium, 
roughly follow the response function predictions, but with quite large 
scatter and systematic offsets from the one-to-one relation. The variations 
in Na, sensitive to sodium are qualitatively well predicted by the response 
functions, for all three star types, but with larger scatter than expected 
from Lick index uncertainties, for the cool stars.

\begin{table*}
 \centering
 \begin{minipage}{175mm}
\begin{center}
  \caption{Reduced chi-squared ($\chi_{\nu}^2$) values for
comparisons of normalised observations (from MILES) versus normalised models 
(using response functions for [Z/H] and [$\alpha$/Fe] changes in stars). 
These $\chi_{\nu}^2$ values take into account errors added in quadrature 
from observational errors in Lick indices and systematic 
offsets due partly to slight base star mismatches. Results are shown for 
three star types (cool giant = CG, turn-off star = TO, cool dwarf = CD)
and for star response functions from Korn et al. 2005 (K05) and 
Houdasheldt et al. 2002 (H02). The final column shows results comparing 
normalised observations versus normalised model, using star response 
functions for [$\alpha$/Fe] changes only, as used in L09 (labelled $\Delta$W12 
here), for 71 stars in MILES.}
  \begin{tabular}{llrrrrrrr}
  \hline
   Index & Index   & \multicolumn{6}{c}{$\chi_{\nu}^2$}  \\
  number & name    & \multicolumn{3}{c}{{\bf ---------K05---------}} 
& \multicolumn{3}{c}{{\bf ---------H02---------}} & {\bf $\Delta$W12 only} \\
         &         & {\bf CG} & {\bf TO} & {\bf CD}   
& {\bf CG} & {\bf TO} & {\bf CD} & {\bf (Any star type)} \\
 \hline
  1 & H$\delta_A$ &   4.61 &   1.75 &  10.57 & 17.81 &  1.76 &  6.71 &  5.09 \\
  2 & H$\delta_F$ &  11.04 &   1.70 &   7.01 &  2.99 &  1.62 &  4.06 &  1.78 \\
  3 & CN$_1$      &   7.96 &   1.02 &   7.52 &  8.78 &  0.61 &  1.15 & 14.58 \\
  4 & CN$_2$      &   5.61 &   0.31 &   1.13 &  7.75 &  0.23 &  1.12 & 11.14 \\
  5 & Ca4227      &   6.45 &   0.70 &   2.06 &  7.79 &  0.49 &  4.55 &  2.64 \\
  6 & G4300       &   0.79 &   2.43 &   0.67 &  1.23 &  2.09 &  0.68 &  4.39 \\
  7 & H$\gamma_A$ &   8.18 &   3.96 &   5.88 &  0.48 &  2.23 &  1.40 &  9.92 \\
  8 & H$\gamma_F$ &   5.76 &   1.87 &   5.61 &  2.45 &  1.92 &  3.36 &  9.08 \\
  9 & Fe4383      &   1.29 &   3.07 &   7.05 &  1.77 &  1.60 &  1.51 &  3.23 \\
 10 & Ca4455      &   1.69 &   0.07 &   0.42 &  0.77 &  0.07 &  0.30 &  0.74 \\
 11 & Fe4531      &   1.04 &   0.52 &   1.53 &  2.00 &  0.99 &  1.61 &  0.76 \\
 12 & C$_2$4668   &  25.13 &   5.75 &   1.32 & 26.35 &  2.45 &  1.58 &  6.60 \\
 13 & H$\beta$    &   1.12 &   2.25 &   1.69 &  1.66 &  1.65 &  1.28 &  1.32 \\
 14 & Fe5015      &   1.19 &   1.49 &   1.52 &  1.14 &  1.49 &  6.68 &  3.02 \\
 15 & Mg$_1$      &   2.06 &   4.80 &   3.00 &  1.69 &  1.12 &  9.27 & 16.46 \\
 16 & Mg$_2$      &   1.94 &   1.70 &  12.07 &  2.23 &  1.28 & 12.69 & 32.91 \\
 17 & Mgb         &   2.73 &   1.38 &   1.39 &  4.04 &  1.18 &  1.43 &  8.16 \\
 18 & Fe5270      &   3.61 &   0.58 &   2.97 &  2.38 &  0.60 &  1.61 &  4.66 \\
 19 & Fe5335      &   1.47 &   0.62 &   4.34 &  2.38 &  0.50 &  4.89 &  3.54 \\
 20 & Fe5406      &   2.09 &   0.35 &   3.44 &  1.90 &  0.34 &  2.84 &  3.24 \\
 21 & Fe5709      &   0.47 &   0.34 &   0.61 &  0.42 &  0.36 &  0.60 &  0.52 \\
 22 & Fe5782      &   0.87 &   0.11 &   0.31 &  0.75 &  0.11 &  0.32 &  0.52 \\
 23 & NaD         &   3.91 &   0.47 &  11.32 &  4.40 &  0.49 & 17.85 &  3.80 \\
\hline
\end{tabular}
\end{center}
\end{minipage}
\label{chisq_results}
\end{table*}

The above results do not change significantly if different base stars 
are used, provided the base stars have the correct atmospheric parameters, 
within observational errors. This helps to confirm that the trends found 
are not just a result of small systematic differences in temperature scales 
between the theoretical and observed stars being compared. 
For the Mg indices (Mg$_1$, Mg$_2$, Mgb) and 
for Ca4227, in CG stars, differential index changes are slightly more 
affected by the choice of base star than in most cases. That is, the 
red circles in Figs.~1 shift significantly, compared to observational 
index errors, with a change in base star (systematically by 
$\sim\pm$0.03 mag for Mg$_1$ and Mg$_2$, and by $\sim\pm$10\% 
for Mgb and Ca4227). Therefore for these indices it is harder to 
accurately check the response function predictions with the empirical 
observations.

A set of response functions for solar abundance models at the younger age of
1 Gyr are given in tables 15 to 17 of K05. Amongst the MILES stars there are 
base stars that match the CD and CG model stars in these tables (but not for 
the TO model star). Therefore, comparisons were made of observed versus 
theoretical normalised indices for these younger CD and CG star models. 
Results showed similar trends and scatters as previously found for the 5 Gyr 
models, but with a slight improvement in the H$\delta_A$ and H$\gamma_A$
indices, for which the normalised observations versus models were closer to
one-to-one trends. Scatter for the H$\gamma$ and H$\delta$ features increased
in the comparisons for the younger age case.

In summary, most Lick indices follow the predictions of the K05 response 
functions as well as we can tell from the empirical data, except for 
H$\gamma_A$, H$\gamma_F$, H$\delta_A$, H$\delta_F$ indices, which 
show systematic deviations from the predictions. These indices lie in 
the blue part of the spectrum where the flux from cool stars is rapidly 
changing with wavelength and where the influence of abundance 
effects is large (see Section 4). Similar results were found
for two sets of MILES stars representing ages of 1 and 5 Gyrs respectively.
Mg$_1$ and C$_2$4668 indices also show systematic deviations from the 
response function predictions in the case of warm TO stars, as described 
above. 

\begin{figure}
 \centering
 \includegraphics[width=84mm, angle=0]{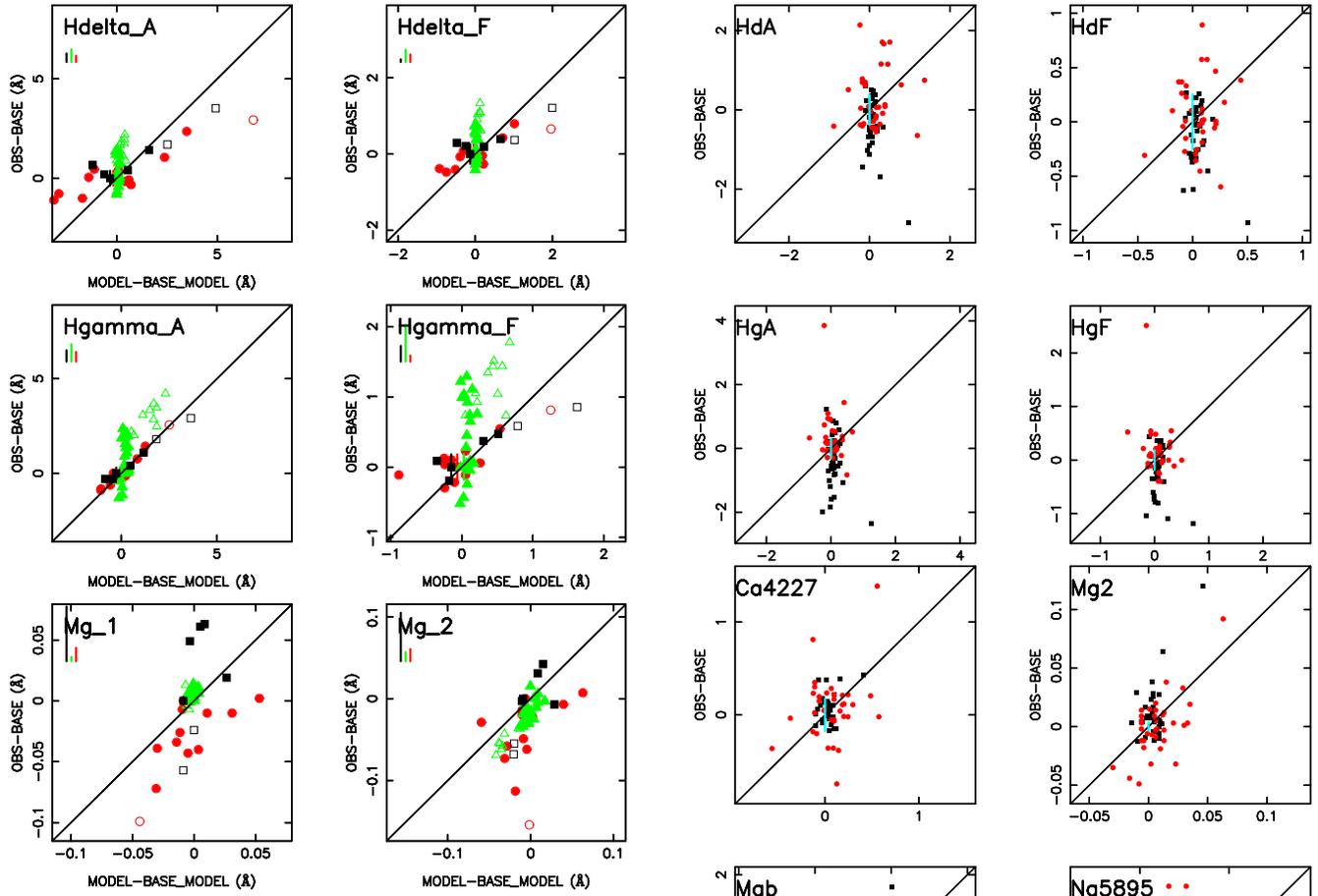} 
 \caption{Testing the response functions of H02. 
Comparison of normalised empirical versus normalised theoretical line 
strengths for standard Lick indices sensitive to H$\gamma$ and H$\delta$ 
Balmer lines, showing similar behaviour to that seen in Fig.~1(b); 
plus Mg$_1$ and Mg$_2$ indices. Symbols as in Fig.~1(a).}
\end{figure}

\subsection{H02 Results}
An alternative set of models exploring response functions was that of
Houdasheldt et al. 2002, who made their response functions for models 
of 3 stars available at: http://astro.wsu.edu/hclee/HTWB02 (H02). 
The H02 response functions differ in value from those of K05 and 
also the line strengths listed at the base abundances are slightly 
different. Here those tables from H02 are used to test the same 3 stars 
as in tables 12 to 14 of K05
(in terms of their $T_{\rm eff}$ and $\log~g$ parameter and base 
solar abundances).
The same procedure as described in Section 3.1 was applied to test the H02 
response functions. Overall the results when comparing to MILES observations
are very similar to what is found for the K05 response functions, with 
some improvements.
Fig.~2 shows the H$\gamma$ and H$\delta$ features using H02 response 
functions, illustrating qualitatively the same problems as with the K05 
response functions. We note however, that H$\delta_F$, H$\gamma_A$ and 
H$\gamma_F$ features for H02 response functions give better agreement 
with the observations, as seen in Fig.~2 and in Table~4. 
Therefore the use of H02 might be preferred over K05, particularly for the 
higher order Balmer indices.
H$\beta$ has similar scatter for both K05 and H02 cases.   
Table~4 also shows that the TO stars are generally better fit by 
the H02 response functions.

For indices that are treated as positive, but which go slightly negative, 
the application of response functions becomes invalid. This is seen for 
Mg$_1$ in TO stars, for K05 response functions (Fig.~1(d)), and for 
H$\beta$ in CD and CG stars for H02 response functions. For H02 response 
functions the Mg$_1$ index (expressed as a line strength) is positive for 
all 3 star types, hence TO stars 
show negligible variation in Mg$_1$ in the model predictions, in agreement 
with the observations. Plots for Mg$_1$ and Mg$_2$ from H02 are also shown in 
Fig.~2. The plot for Mg$_2$ using the H02 response functions looks 
similar to that in Fig.~1(e) in spread and offsets (also true for Mgb), 
indicating similar results compared with those of the K05 response functions.

\subsection{L09 Results}

Lee et al 2009 (L09) note that, in their models, the broader H$\gamma_A$ and 
H$\delta_A$ Balmer features are significantly affected by iron abundance.
In their on-line comparisons with K05 response functions for individual 
star models, their plot for H$\delta_A$, for example, shows an increase 
of 2\AA\ (or 5.5\AA), for a +0.3 enhancement in [$\alpha$/Fe] at constant 
[Fe/H]=0.0 (or [Z/H]=0.0); see: http://astro.wsu.edu/hclee/HdA.pdf, 
red square (or blue square). It is difficult to compare this directly 
with the spread in observational data for this index, shown in Fig.~1(b),
since those data include stars with a range of both [$\alpha$/Fe] and 
[Fe/H] values, however, the spread in the observations is less than 
$\sim$ 5\AA\ over a broad range in composition, implying that the L09
models may also overestimate the variation expected in this index. 

Although L09 made use of extensions following on from the work of H02,
their on-line plots for individual star model response function behaviour
indicate different values than those in the H02 tables. Therefore it would 
be very interesting to be able to test the star response functions of L09. 
However, L09 did not publish tables of response functions for their 
350 model stars (only for their SSPs). 
Use of the L09 SSP response functions (their table~5) therefore 
relies on the assumption that they have included different phases of 
stellar evolution in their correct proportions. It is likely to be better 
than the use of only 3 stars as is often done in determining response 
functions for SSPs, however, their published data do not allow us 
to test the understanding on a star-by-star basis, as we are attempting 
in this paper. 

\begin{figure}
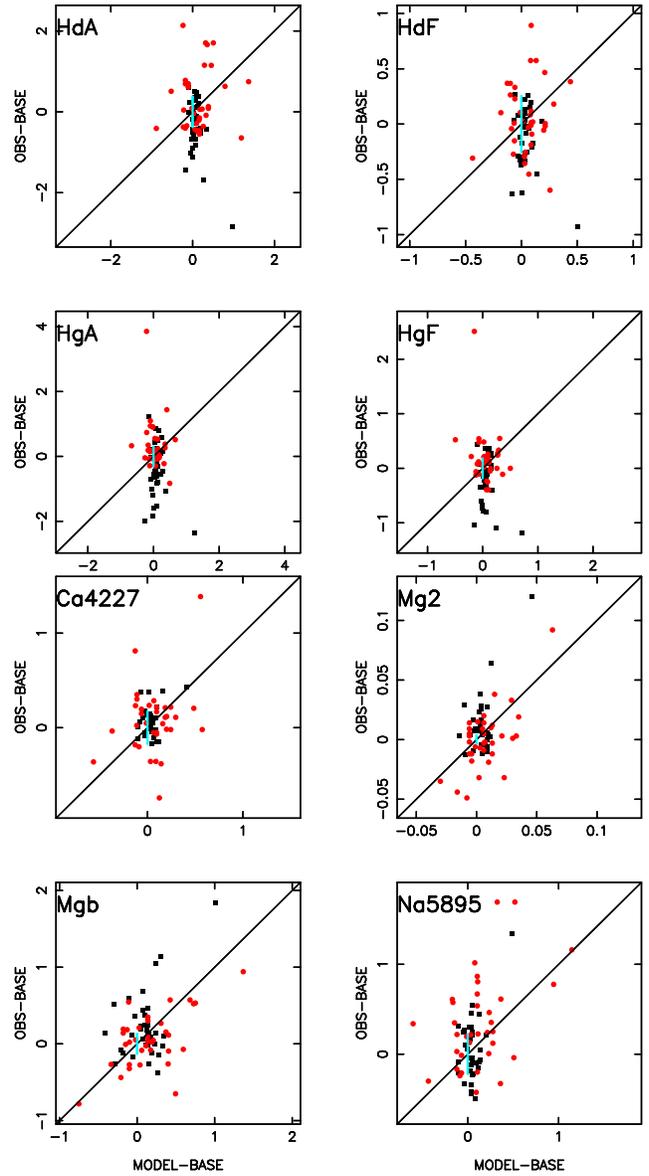

 \includegraphics[width=84mm, angle=0]{pgpage1W12_71stars_HgHd_square_noXlabel.eps}
 \includegraphics[width=84mm, angle=0]{pgpage1W12_71stars_CaMgNa_square.eps}
 \caption{Testing the response functions of W12
for the effects of [$\alpha$/Fe] changes only. 
Comparison of empirical versus theoretical line-strength differences
for standard Lick indices in individual stars. 
The empirical observations are for stars in the MILES library, 
with know [Fe/H] and [$\alpha$/Fe] (from [Mg/Fe] measurements in M11).
This plot excludes open cluster stars and one star with $|$[Mg/Fe]$|>0.25$.
Two star types are shown: cool dwarfs (CD, black squares), 
and cool giants (CG, red circles), divided at $\log~g=3.0$. 
The observed stars shown are chosen to 
all have the same $T_{\rm eff}$, $\log~g$ and [Fe/H] as the base stars within 
observational errors. The base stars used are from MILES and have 
[Mg/Fe]=0.0, within a small tolerance of $\pm$0.01 dex. 
Average observational errors (2$\sigma$) are attached to one of the 
star points for each index. The straight line shows the one-to-one 
relation in each case. Both empirical (vertical) and theoretical (horizontal) 
axes show {\it differences} in Lick indices between observed stars and their
matching base stars.}
\end{figure}

Worthey (priv. comm.- hereafter W12) provided us with model response functions 
for stars and software used in L09 to generate index changes due to chemistry. 
These allowed the present authors to generate changes in indices for stars 
of user defined atmospheric parameters. This information was used to derive 
three tables equivalent to tables 12 to 14 in K05, for changes 
due to individual elements in CD, TO and RG stars. For overall metallicity 
the changes of K05 were assumed, since overall metallicity changes were not 
available in the W12 star response functions.
Using these W12 response functions led to similar results as found using 
the K05 response functions shown in Fig.~1(a-e), for individual stars. 
The discrepancies in H$\gamma$ and H$\delta$ indices remained.
This similarity of results, using K05 overall metallicity changes with W12
changes to individual elements, supports the fact that overall metallicity 
is the dominant effect for most indices. That is, we are not finding different 
results using W12 changes to individual elements. 

To probe the effects of [$\alpha$/Fe] changes only, the W12 response functions 
were used as follows. Stars with [Mg/Fe] close to 0.0 ($\pm$0.01) in MILES were 
selected, providing 33 base stars.
Matching stars in MILES with the same $T_{\rm eff}$, $\log~g$, [Fe/H] as these, 
within errors, but with differing [Mg/Fe] gave 80 matches. Of these, 
8 were associated with star clusters and were removed since they were 
generally lower signal-to-noise.
One other star with large [Mg/Fe]=+0.454 was also removed to avoid 
large changes in metallicity. The remaining stars
all had $|$[Mg/Fe]$|<0.25$. This latter 
restriction is applied here because these W12 response functions
do not allow us to track the effects of overall metallicity changes, 
therefore we can only use them to test trace element changes.
Observed index differences for the remaining 71 stars
were compared with index differences predicted 
from W12 models. Fig.~3 shows the results for dwarfs stars (black
squares, $\log~g\ge3.0$) and giant stars (red circles, $\log~g<3.0$).
This shows data points scattering about the one-to-one line for each index,
with little sign of any correlations except in the case of magnesium 
and sodium (Mg$_2$, Mgb, NaD), for which the correlation coefficients 
are 0.57, 0.48 and 0.43 respectively. These are significant at $>$99.9\% 
confidence levels, for 71 data points. For these few indices there 
is evidence
that observed indices changes broadly follow indices changes predicted
by W12 response functions. The H$\gamma$ and H$\delta$ features show
somewhat larger scatter than expected from the typical observational
errors (particularly for the giant stars), but no systematic effects 
are evident that would imply an [$\alpha$/Fe] dependence that is 
different between the observations and models. Mean offsets 
are all $<$0.2 dex. Some of the scatter in 
these differential [$\alpha$/Fe] changes may be due to the fact that the
W12 response functions are evaluated from a specified [Fe/H] but keep 
overall metallicity constant; whereas for the observed stars, their 
metallicity is characterised by [Fe/H] so we are not exactly comparing 
like with like, especially at increasingly non-solar [$\alpha$/Fe]. 
Table~4 shows that the CN, H$\gamma$ and Mg sensitive features
agree least within the observational errors for these differential changes.
Thus the W12 response functions may be most useful for modelling 
the effects of element abundance changes, when they can be treated as 
trace element abundances changes. However, for the current comparisons, 
this runs into the finite errors on the [Mg/Fe] measurements, therefore 
weakening this test of the W12 response functions. More accurate measurements 
of abundances and responses of indices to overall metallicity would be
needed to better test the response functions of W12, used in L09.

In summary, these results indicate that the systematic deviations seen 
in MILES observations relative to K05, for the H$\gamma$ 
and H$\delta$ features, may result from insufficiently accurate 
accounting for effects of overall metallicity changes in those response 
functions. The response functions of H02 agree slightly better
with observations for those features. However, they 
are only available for three model stars. The star response functions of 
W12 (priv. comm.) provide the widest scope for testing trace element 
abundance changes but do not allow changes in overall metallicity to be
easily tested. Therefore there is as yet no one set of response functions 
that provide the widest and best fitting to star data. Caution should be
exercised particularly in interpreting the H$\gamma$ and H$\delta$ indices
in stellar populations, plus indices that reach values close to zero 
in some stars (H$\beta$, Mg$_1$) when using response functions.

\section[]{Comparisons of Spectra}

Next we go on to test attributes of spectra (rather than indices) to 
varying abundance patterns.

\subsection{Comparison with published model spectra}

\citet{b20001}
were the first to compare their theoretical model spectra for 
enhanced and unenhanced stars. They found the largest differences 
in the blue part the spectrum, particularly when comparing at 
constant overall metallicity.

In the current analysis, ratios were created for models of star spectra 
published by \citet{b501},
for a typical dwarf star ($T_{\rm eff}$=5500K, $\log~g=4.0$, [Z/H]=-0.2) and a 
typical giant star ($T_{\rm eff}$=4500K, $\log~g=2.0$, [Z/H]=-0.2) for enhanced 
([$\alpha$/Fe]=+0.4) over un-enhanced ([$\alpha$/Fe]=0.0) models, where 
$\alpha$-elements are considered to be 
O, Ne, Mg, Si, S, Ca and Ti (see Coelho et al. 2005, section 3.1). The 
theoretical spectra are published at fixed [Fe/H] values and at these two 
[$\alpha$/Fe] values. These spectra were interpolated to obtain spectra with 
overall metallicities at sub-solar abundance [Z/H]=-0.2, using the 
transformations given by \citet{b5} (their table~1). 
This value of overall metallicity was chosen in order to maximize 
the possibility of finding similar enhanced and unenhanced stars 
in the observations. Hence the model spectral ratios are compared 
in Fig.~4 with similar ratios, made from interpolating empirical 
dwarf star spectra in the MILES library, for particular values 
of [Fe/H] amd [Mg/Fe]. The interpolator used is 
an extended version of the 3-dimensional interpolator described in 
\citet{b303} and \citet{b301}
that now allows the user to select stars by [Mg/Fe] (from M11), as a proxy 
for [$\alpha$/Fe], within the limits imposed by the MILES library coverage 
of this parameter. This also approximates the link from [Fe/H] to an 
estimate of overall metallicity [Z/H] by assuming the transformation
given by \citet{b5} (their table~1).
Empirical spectra used in Fig.~4 are approximate in enhanced 
[$\alpha$/Fe] values, due to the limited range of such stars available 
in the local Solar neighbourhood (as can be seen in fig.~10 of M11).

Qualitatively we see a good agreement between theoretical and empirical 
spectral ratios plotted in Fig.~4. There are some differences in detail, 
particularly in the complex spectral region blue-ward of about 4500\AA, 
which are likely to be at least partialy attributed to differences in 
C, N and O abundances between theoretical models and MILES stars. 
There are the features modelled by \citet{b22}
that affect this region, including CNO3862, CNO4175 
as well as the CN bands and features due to other elements.
New theoretical spectral models are currently being generated 
(Coelho - private communication) and a more quantitative comparison will 
await those models.

\begin{figure*} 
\begin{center} 
\includegraphics[width=80mm, angle=-90]{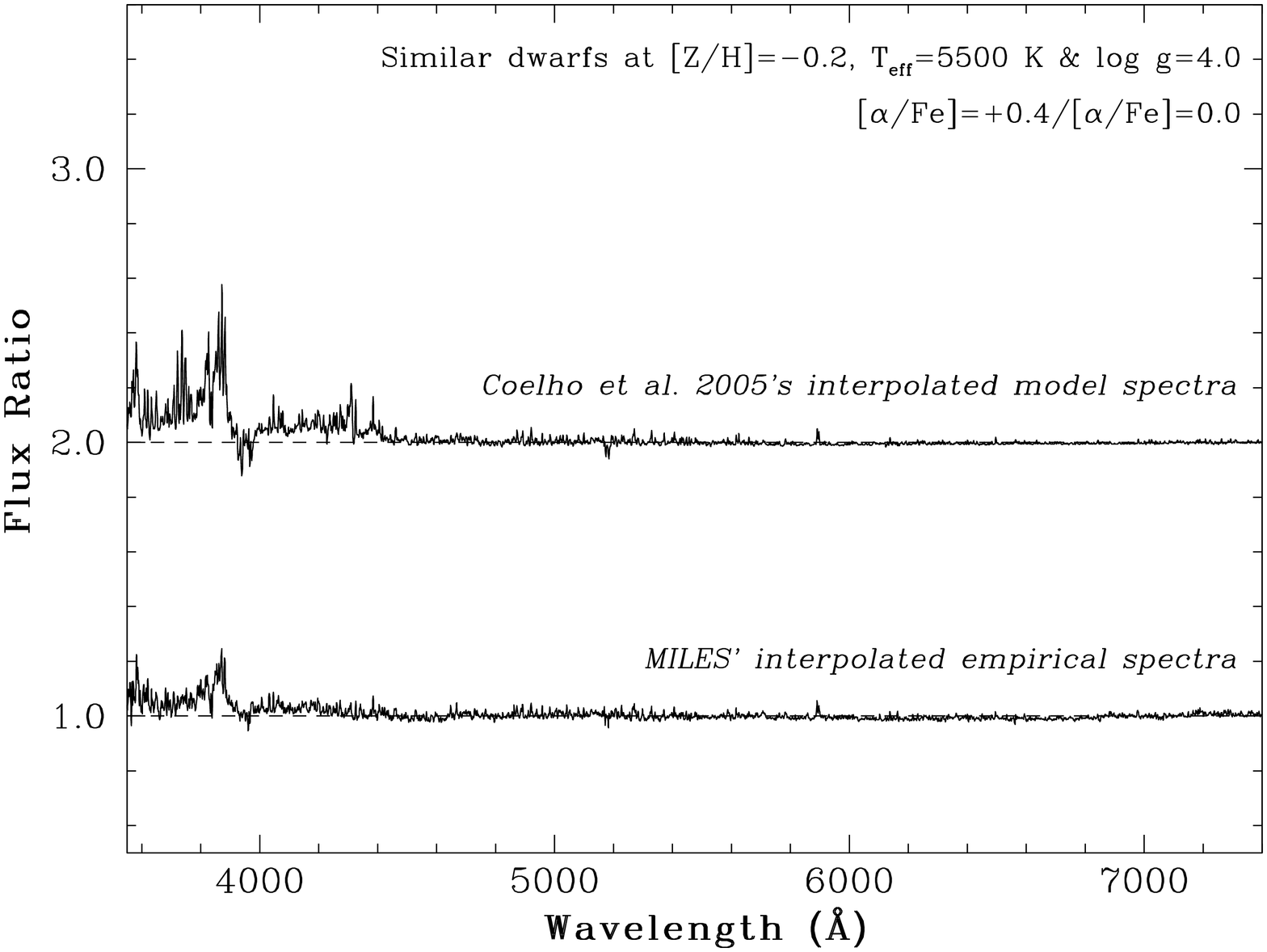}
\includegraphics[width=80mm, angle=-90]{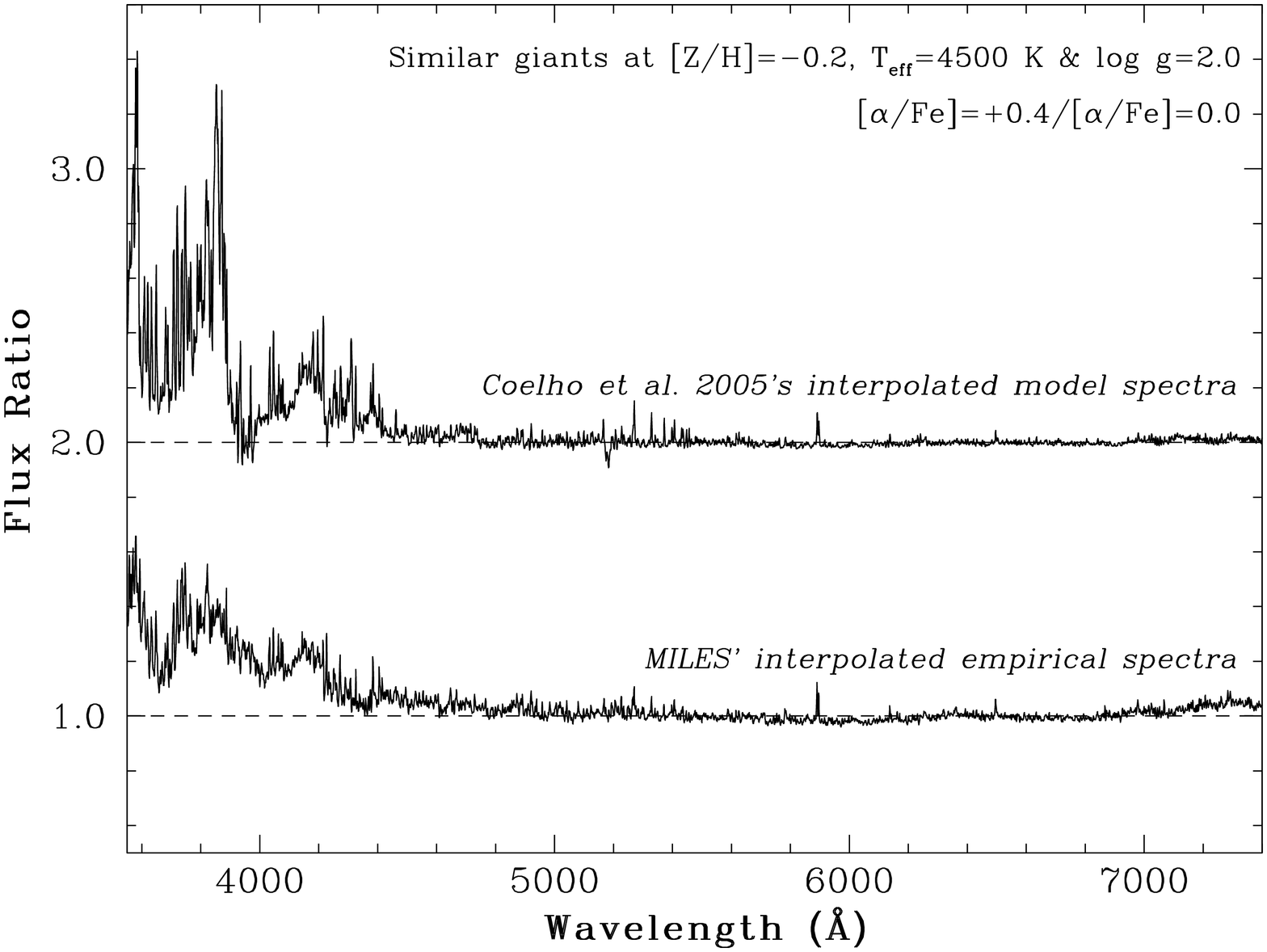} 
\end{center} 
\caption{Comparison of ratios of spectra with enhanced 
([$\alpha$/Fe]=+0.4, [Fe/H]=-0.5) and un-enhanced ([$\alpha$/Fe]=0.0, 
[Fe/H]=-0.2) abundance patterns, corresponding to fixed overall 
metallicity ([Z/H]=-0.2) in the models. Enhaced spectra are divided
by un-enhanced spectra in each case. The top plot shows dwarf stars 
($T_{\rm eff}$=5500K, $\log~g=4.0$). The lower plot shows giant stars 
($T_{\rm eff}$=4500K, $\log~g=2.0$). In each plot: the upper spectral ratio is 
theoretical, obtained by interpolating theoretical spectra from the 
library of Coelho et al. 2005 and the lower spectral ratio is observational, 
obtained by interpolating the MILES empirical spectral library. The theoretical
spectral ratios are vertically offset by 1.0 to separate them from the observed 
spectral ratios shown.}  
\end{figure*}

\subsection{Comparisons of empirical spectra for specific stars}

To qualitatively investigate the influence of $\alpha$-element abundance 
on empirical stellar spectra, stars were chosen in pairs with similar 
photospheric parameters in the MILES [Mg/Fe] Catalogue (M11), for a few 
representative evolutionary stages in the context of SSP modelling.
The evolutionary stages analysed are: red normal giant with 
$T_{\rm eff}$ $\approx$ 4000K and $\log~g$ $\approx$ 1.5 
(around K5III), main sequence turn-off star with 
$T_{\rm eff}$ $\approx$ 6600K and $\log~g$ $\approx$ 4.2 (around F4V)
for an SSP of about 4 Gyr, and a cool main sequence dwarf with 
$T_{\rm eff}$ $\approx$ 5100K and $log~g$ $\approx$ 4.5
(around K1V).

The basic approach was to compute divisions of spectra
by choosing pairs of similar stars in terms of $T_{\rm eff}$ and $\log~g$
with different abundances, keeping either [Z/H], [Fe/H] or [Mg/H] constant 
within some level. We assumed the solar abundance pattern from \citet{b54},
as adopted by \citet{b501}.
We calculated overall [Z/H] from an abundance pattern, 
generating values for various combinations of [Fe/H] and [$\alpha$/Fe], 
assuming all alpha elements are elevated to the same level. Fitting a 
bi-variable linear function to the results gives the following 
transformation: 

\begin{equation}
[Z/H] = [Fe/H] + 0.75(\pm 0.02) [\alpha/Fe] + 0.007(\pm 0.006)
\end{equation}

\noindent valid over the ranges: $-1.5\le$[Fe/H]$\le+0.3$ and 
$-0.2\le$[$\alpha$/Fe]$\le+0.6$, and accurate over this range to within 
$<\pm$0.01 dex ($rms$). This fitted equation gives very similar results to 
the correspondences tabulated by \citet{b5}, their table~1,
which also assumed that all $\alpha$ elements varied in the same way.
We used the relation in Equation 6 to search for pairs of stars in 
MILES [Mg/Fe] catalogue (M11), assuming [$\alpha$/Fe] = [Mg/Fe].

Firstly, assuming [Z/H] constant (but [Fe/H] and [Mg/H] varying),
spectra of MILES stars with larger and smaller values of [Mg/H] were divided.
Then, considering [Fe/H] unchanged (but [Z/H] and [Mg/H] varying),
ratios of spectra of MILES stars were computed, with larger divided by 
smaller [Mg/H]. To specifically evaluate the influence of [Fe/H] variation 
on the spectrum as well, ratios of spectra were also obtained by changing 
[Fe/H] (and [Z/H]) with [Mg/H] constant. The spectrum of a MILES star with 
larger [Mg/Fe] (and smaller [Fe/H])
was divided by the spectrum of its analogue with smaller 
[Mg/Fe].

In this approach, any quantitative change in [Mg/H], [Fe/H] or [Z/H] 
for each comparison needs to be taken into account for a more precise 
analysis of the results. The differential relationship among the metal 
abundance parameters is
d[Z/H] = d[Fe/H] + 0.75 d[$\alpha$/Fe]
, or:

\begin{equation}
d[Z/H] = 0.25 d[Fe/H] + 0.75 d[\alpha/H]
\end{equation}

in order to express all parameters on a scale relative to hydrogen.

Table~5
presents the set of star pairs adopted for each of 
the three evolutionary stages considered in the current spectral comparisons.
We searched for pairs of similar stars in the MILES [Mg/Fe] Catalogue 
with differences in $T_{\rm eff}$ and $\log~g$ less than or 
equal to 50K and 0.1 respectively. These fiducial values represent 
half of one standard deviation of the temperature and gravity errors 
for FGK stars in the MILES library \citep{b2}.
A very restrictive condition to fix [Z/H], [Fe/H] or [Mg/H] has also 
been imposed so that the maximum difference in each metal abundance 
parameter ($\Delta([X/H])_{max}$ 
is assumed to be $\le 0.05$ dex for each pair of stars.
Gravity differences had to be somewhat relaxed in order to find some
suitable pairs of stars, such that $\Delta\log~g\le0.2$ for the red giants
with fixed [Fe/H]; $\le0.5$ for cool dwarfs with [Fe/H] fixed and $\le0.3$
for all turn-off stars. The abundance similarities also needed to be 
relaxed to $\Delta[X/H]\le0.07$ for cool dwarfs.

\subsubsection{Normal red giant stage}
Considering [Z/H] fixed around the solar value, the more magnesium-enhanced 
red normal giant presents a flux excess in the blue part of spectrum 
(see Fig.~5a). This excess is a result of the increasing of 
[Mg/H] and/or decreasing [Fe/H]. When [Fe/H] is 
assumed constant, with [Z/H] changing below the solar value (Fig.~5b) however, 
the spectral ratio is relatively flat. By keeping [Mg/H] unchanged close to 
solar abundance and varying [Fe/H], the spectrum ratio 
shows that the excess in the blue flux is due to a smaller [Fe/H] (see Fig.~5c).
The conclusion is that the blue flux excess for 
$\alpha$-enhanced red giants in comparison with less $\alpha$-enriched ones
at a fixed overall metallicity [Z/H] occurs mainly due to a decrease in [Fe/H] 
instead of an increase in [$\alpha$/Fe]. However, the level of this effect is 
somewhat uncertain in the data since the example shown uses an Algol-like 
system (HD192909).
There is some limited capacity to check this result with other pairs 
of similarly cool red giant stars in MILES. The spectral ratios found vary, 
but qualitatively show the same results in most cases. 
Hotter red giant stars show less variation (also see Section 5). 
This analysis is also limited by the observational errors on all
photosperic parameters involved, as qualitatively stressed 
for temperature and gravity, later in this Section. 
Therefore we have attempted to concentrate on the most reliable cases.

\subsubsection{Main sequence turn-off dwarf for an evolved SSP}
Following a similar procedure for three pairs of turn-off stars showed 
that they vary much less with abundance changes, but still varying most 
in the blue. Results are shown in Fig.~6, with a smaller vertical 
scale than used for the red giants in Fig.~5. This smaller variation due 
to chemistry is not so surprising since these are hotter stars (see Section 5).
A few other similar pairs of stars show qualitatively similar behaviour.

\subsubsection{Cool main sequence dwarf}
Spectral ratios for three pairs of cool dwarf stars are plotted in Fig.~7 with 
the same vertical scale as for the turn-off stars in Fig.~6. Variations are 
smaller than for the cool red giant stars, but greater than for the warm 
turn-off stars. When [Fe/H] is kept fixed (central plot in Fig.~7) there 
are least variations in the blue. When [Mg/H] is kept fixed (lowest plot 
in Fig.~7) there are small variations, particularly in the blue, mainly due to
changes in [Fe/H]. A few similar examples can be found in the MILES data, 
qualitatively supporting the relative behaviour shown in Fig.~7. 

In future, these spectral ratios will be compared with their exact 
counterparts in the new theoretical models currently being generated.

\begin{table*} 

 \centering
 \begin{minipage}{175mm}

\caption{
Set of three pairs of similar MILES stars for a) red giant branch (RGB) stage, 
b) turn-off (TO) stage and c) cool dwarf (CD) stage. The first two rows in each 
of these evolutionary categories show the photospheric parameters of a pair of 
similar stars with fixed [Z/H] around the solar value but varying [Mg/Fe], [Mg/H] 
and [Fe/H] ($\mid\Delta$([Mg/H])$\mid$ $\geq$ 0.1 dex).
The intermediate two rows present the stellar parameters for a pair of similar
stars with [Fe/H] fixed around the solar value but changing [Mg/Fe], [Mg/H] and 
[Z/H] ($\mid\Delta$([Mg/H])$\mid$ $\geq$ 0.25 dex).
In the last two rows the parameters of another pair of similar stars is shown
with [Mg/H] constant but varying [Mg/Fe], [Fe/H] and [Z/H] 
($\mid\Delta$([Fe/H])$\mid$ $\geq$ 0.25 dex). Parameters are from the MILES
library, except that [Mg/H], [Mg/Fe], $\sigma$[Mg/Fe] and Notes are from M11, 
plus [Z/H] is from equation 6. The final column indicates whether [Mg/Fe] is 
from medium or high resolution spectral studies (see M11 for further details).}
\label{RGB_spectral_comparison} 
\begin{center} 
\begin{tabular}{@{}lllllrrrrrl} 
\hline 
\hline 
\#MILES     & Type   & Name            & $T_{\rm eff}$ & $\log~g$ & [Fe/H] & [Mg/Fe] & $\sigma$Mg/Fe & [Mg/H]  &[Z/H] &  Cat Notes \\
\hline 
{\bf a) Red Giants}  &   &             & (K)          &         &  (dex) &  (dex)  &  (dex)  &  (dex)  & (dex) &           \\ 
\hline 
\hline 
{\bf [Z/H] constant} &   &             &              &         &        &         &       &          &       &            \\
0760F       & field  & HD192909        &   3880       &  1.34   & -0.43  &  0.53   & 0.15  &  0.10    & -0.03 &  mr Mg5528 \\
0650F       & field  & HD164058        &   3902       &  1.32   & -0.05  &  0.02   & 0.16  & -0.03    & -0.03 &  HR Ae01   \\
\hline 
{\bf [Fe/H] constant} &  &             &              &         &        &         &       &          &       &            \\
0059F       & field  & HD009138        &   4103       &  1.85   & -0.37  &  0.19   & 0.10  & -0.18    & -0.22 &  mr BothMg \\
0557F       & field  & HD137704        &   4109       &  1.97   & -0.37  & -0.16   & 0.13  & -0.53    & -0.48 &  mr Mg5183 \\
\hline 
{\bf [Mg/H] constant} &  &             &              &         &        &         &       &          &       &            \\
0760F       & field  & HD192909        &   3880       &  1.34   & -0.43  &  0.53   & 0.15  &  0.10    & -0.03 &  mr Mg5528 \\
0561F       & field  & HD139669        &   3895       &  1.41   & -0.01  &  0.06   & 0.15  &  0.05    &  0.04 &  mr Mg5528 \\
\hline 
\hline 
{\bf b) Turn-off Stars}  &   &         & (K)          &         &  (dex) &  (dex)  &  (dex)  &  (dex)  & (dex) &           \\ 
\hline 
\hline 
{\bf [Z/H] constant} &   &             &              &         &        &         &       &          &       &            \\
0444F       & field  & HD109443        &   6632       &  4.20   & -0.65  &  0.43   & 0.10  & -0.22    & -0.32 &  mr BothMg \\
0525F       & field  & HD130817        &   6585       &  4.08   & -0.46  &  0.14   & 0.10  & -0.32    & -0.35 &  mr BothMg \\
\hline 
{\bf [Fe/H] constant} &  &             &              &         &        &         &       &          &       &            \\
0482F       & field  & HD119288        &   6594       &  4.03   & -0.46  &  0.53   & 0.10  &  0.07    & -0.06 &  mr BothMg \\
0412F       & field  & HD099747        &   6604       &  4.06   & -0.51  &  0.16   & 0.10  & -0.35    & -0.38 &  mr BothMg \\
\hline 
{\bf [Mg/H] constant} &  &             &              &         &        &         &       &          &       &            \\
0444F       & field  & HD109443        &   6632       &  4.20   & -0.65  &  0.43   & 0.10  & -0.22    & -0.32 &  mr BothMg \\
0504F       & field  & HD125451        &   6669       &  4.44   &  0.05  & -0.22   & 0.10  & -0.17    & -0.11 &  mr BothMg \\
\hline 
\hline 
{\bf c) Cool Dwarfs}  &   &            & (K)          &         &  (dex) &  (dex)  &  (dex)  &  (dex)  & (dex) &           \\ 
\hline 
\hline 
{\bf [Z/H] constant} &   &             &              &         &        &         &       &          &       &            \\
0145F       & field  & HD026965        &   5073       &  4.19   & -0.31  &  0.34   & 0.12  &  0.03    & -0.05 &  HR T98LH05   \\
0684F       & field  & HD171999        &   5031       &  4.65   & -0.10  & -0.03   & 0.15  & -0.13    & -0.12 &  mr Mg5528   \\
\hline 
{\bf [Fe/H] constant} &  &             &              &         &        &         &       &          &       &            \\
0529F       & field  & HD132142        &   5108       &  4.50   & -0.55  &  0.34   & 0.05  & -0.21    & -0.29 &  HR BM05   \\
0138F       & field  & HD025673        &   5150       &  4.50   & -0.60  &  0.07   & 0.05  & -0.53    & -0.54 &  HR BM05   \\
\hline 
{\bf [Mg/H] constant} & &             &               &         &        &         &       &          &       &            \\
0750F       & field  & HD190404        &   5051       &  4.45   & -0.17  &  0.39   & 0.05  &  0.22    &  0.13 &  HR BM05   \\
0322F       & field  & HD075732        &   5079       &  4.48   &  0.16  &  0.09   & 0.05  &  0.25    &  0.23 &  HR BM05   \\
\hline 
\hline 

\end{tabular} 
\end{center} 

\end{minipage}
\label{star_pairs}
\end{table*}

\begin{figure}
\includegraphics[width=67mm, angle=-90]{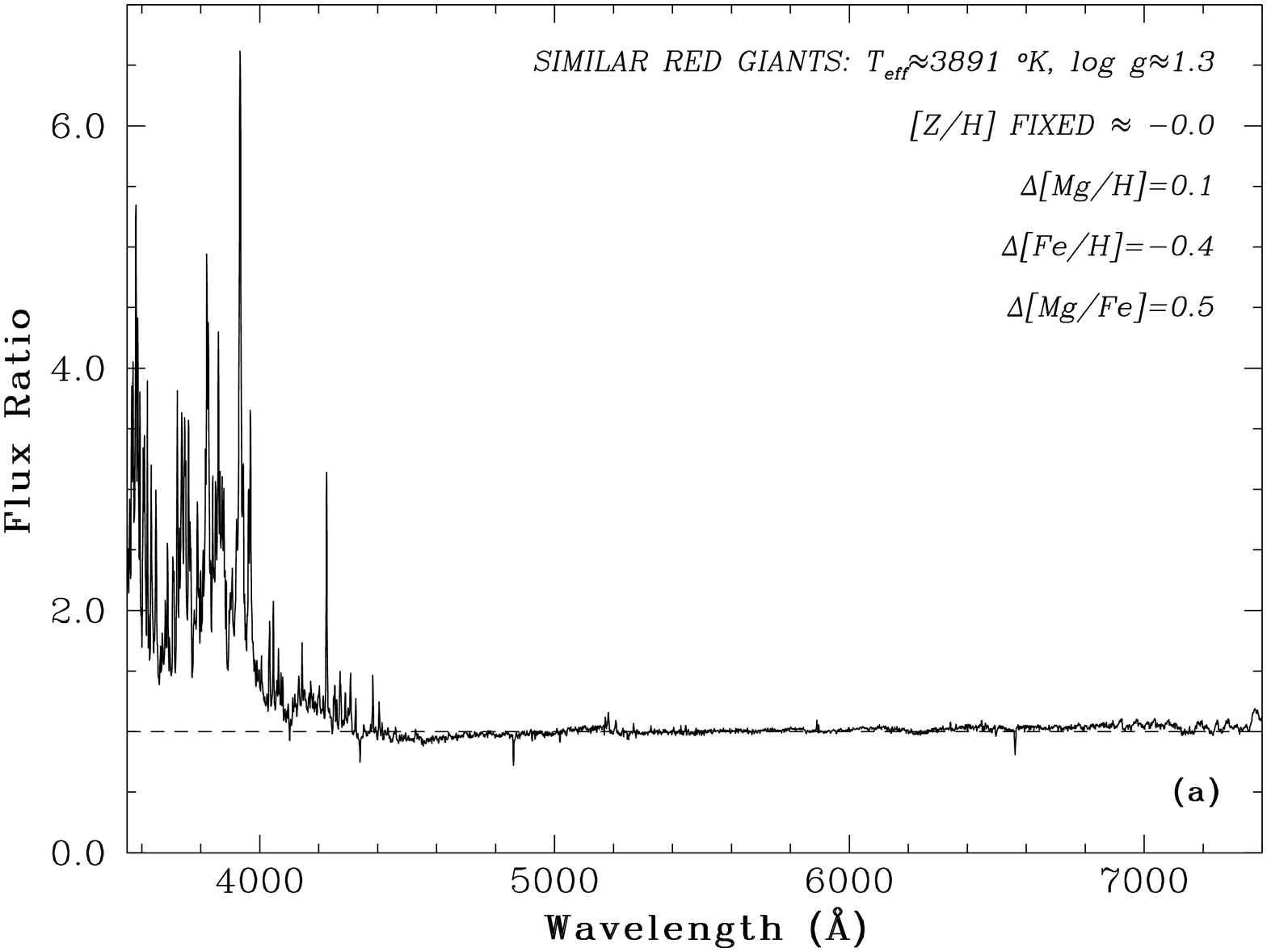}
\includegraphics[width=67mm, angle=-90]{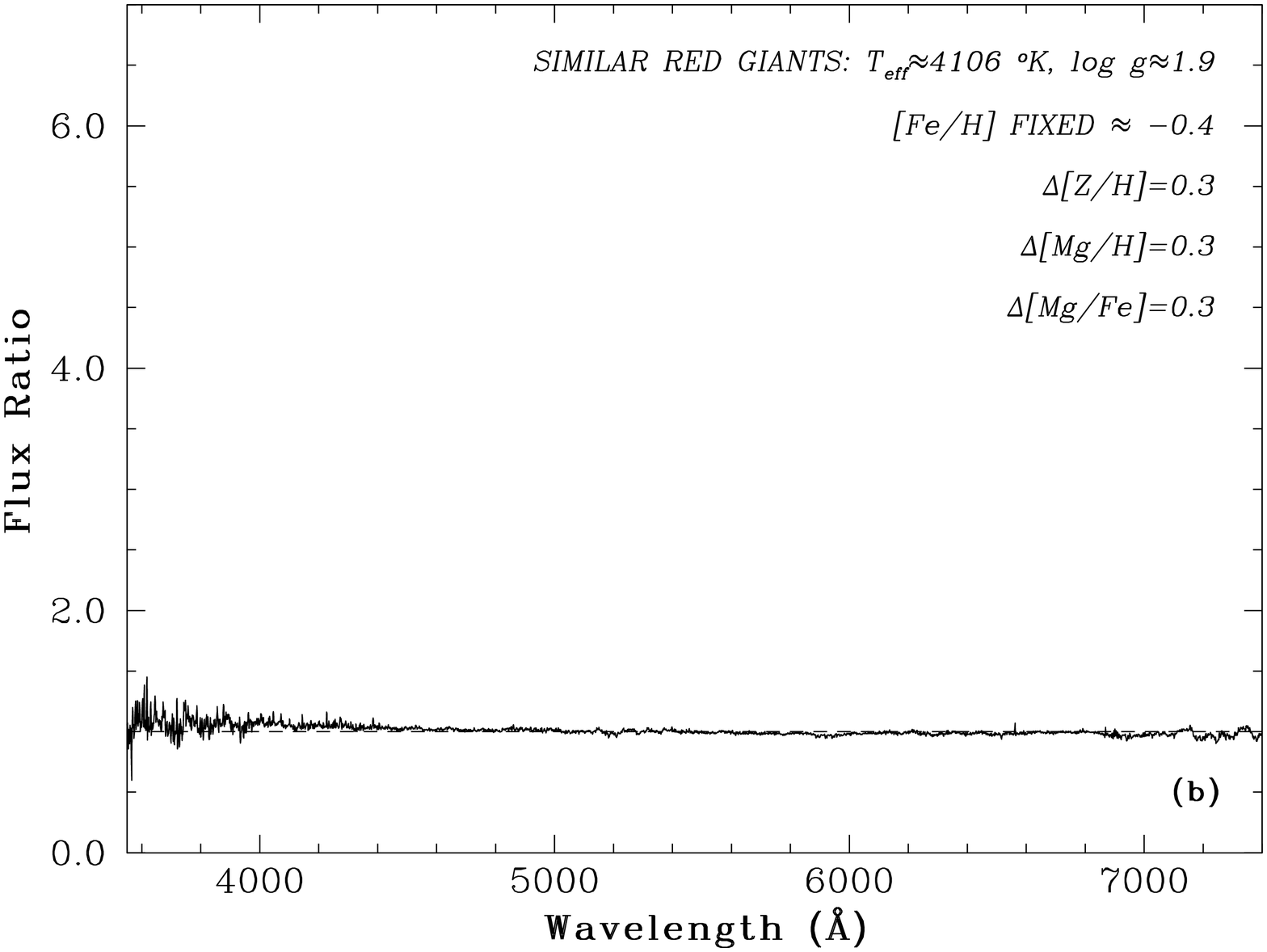} 
\includegraphics[width=67mm, angle=-90]{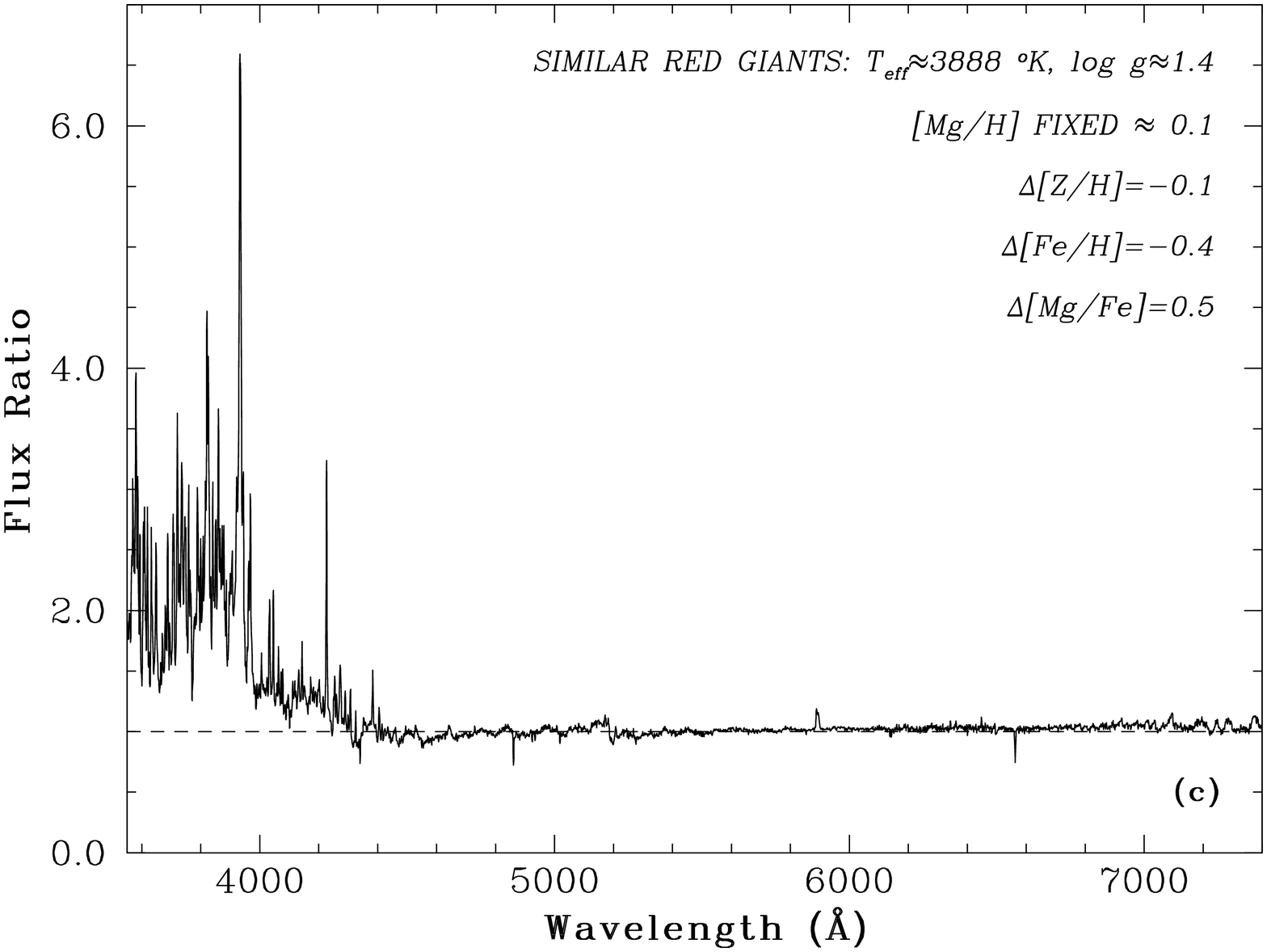}
\caption{ 
Ratios of pairs of MILES spectra  
a) of a pair of analogue stars for the RGB stage, 
fixing [Z/H] around the solar value but varying [Mg/H] and [Fe/H]
($\Delta$([Mg/H]) = +0.13 and $\Delta$([Fe/H]) = -0.38 dex).
b) of a pair of analogue stars for the RGB stage, 
keeping [Fe/H] fixed slightly below the solar value but changing 
[Mg/H] and [Z/H]
($\Delta$([Mg/H]) = +0.35 and $\Delta$([Z/H]) = +0.26 dex).
c) of a pair of analogue stars for the RGB stage, 
assuming [Mg/H] constant but varying [Fe/H] and [Z/H]
($\Delta$([Fe/H]) = -0.42 and $\Delta$([Z/H]) = -0.07 dex).
The star names and their parameters are listed in Table~5.
}
\label{FigX1} 
\end{figure}

\begin{figure}
\includegraphics[width=67mm, angle=-90]{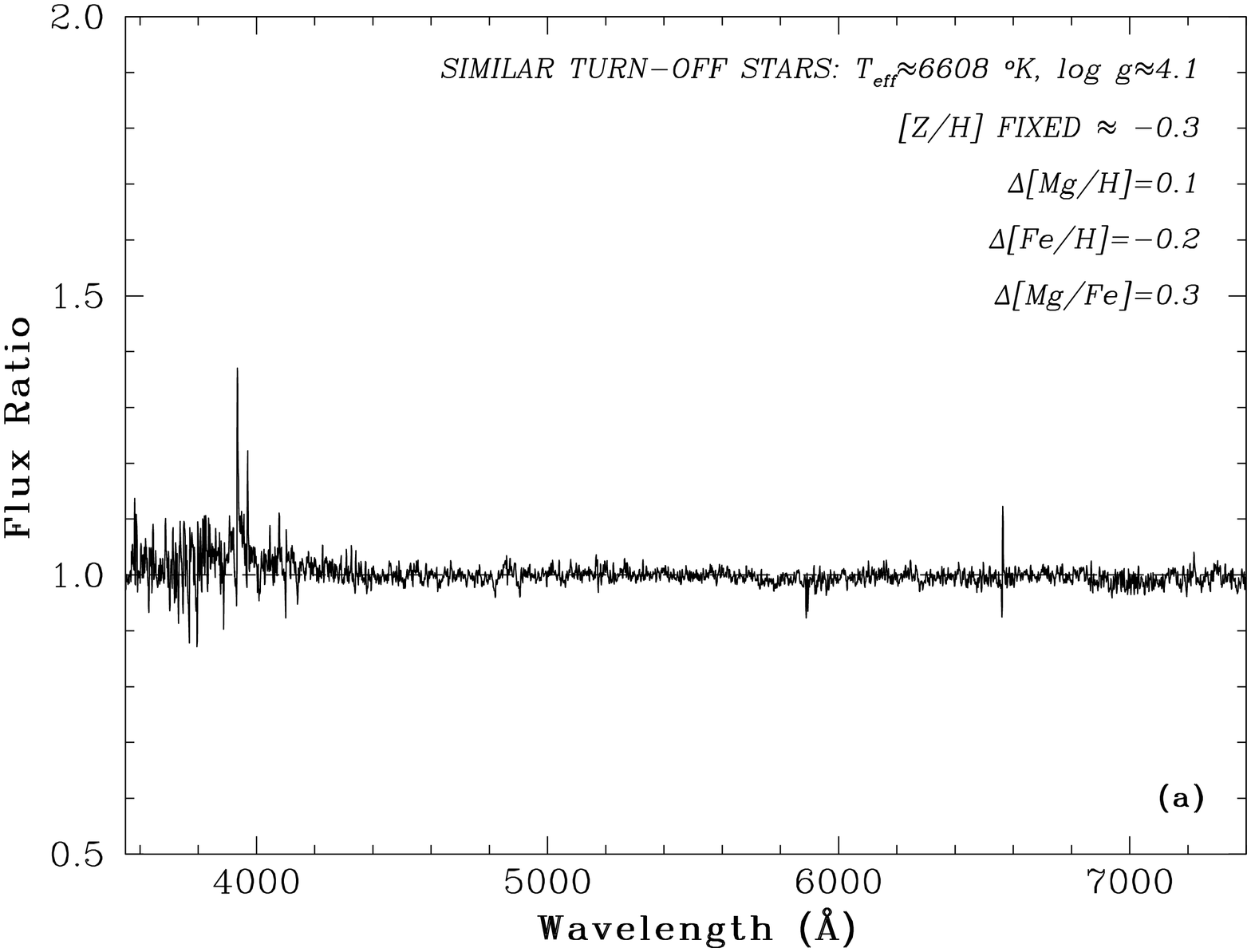}
\includegraphics[width=67mm, angle=-90]{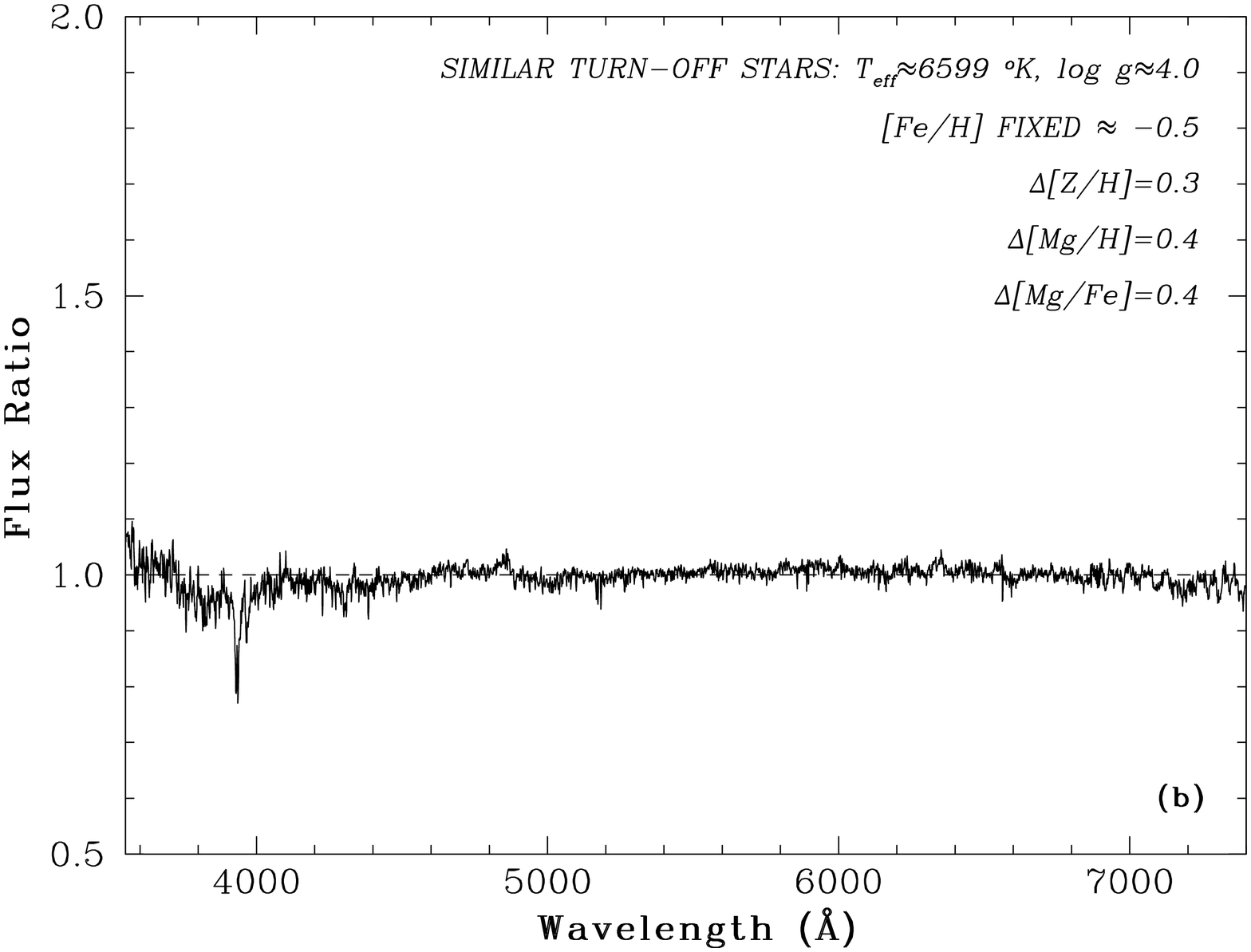} 
\includegraphics[width=67mm, angle=-90]{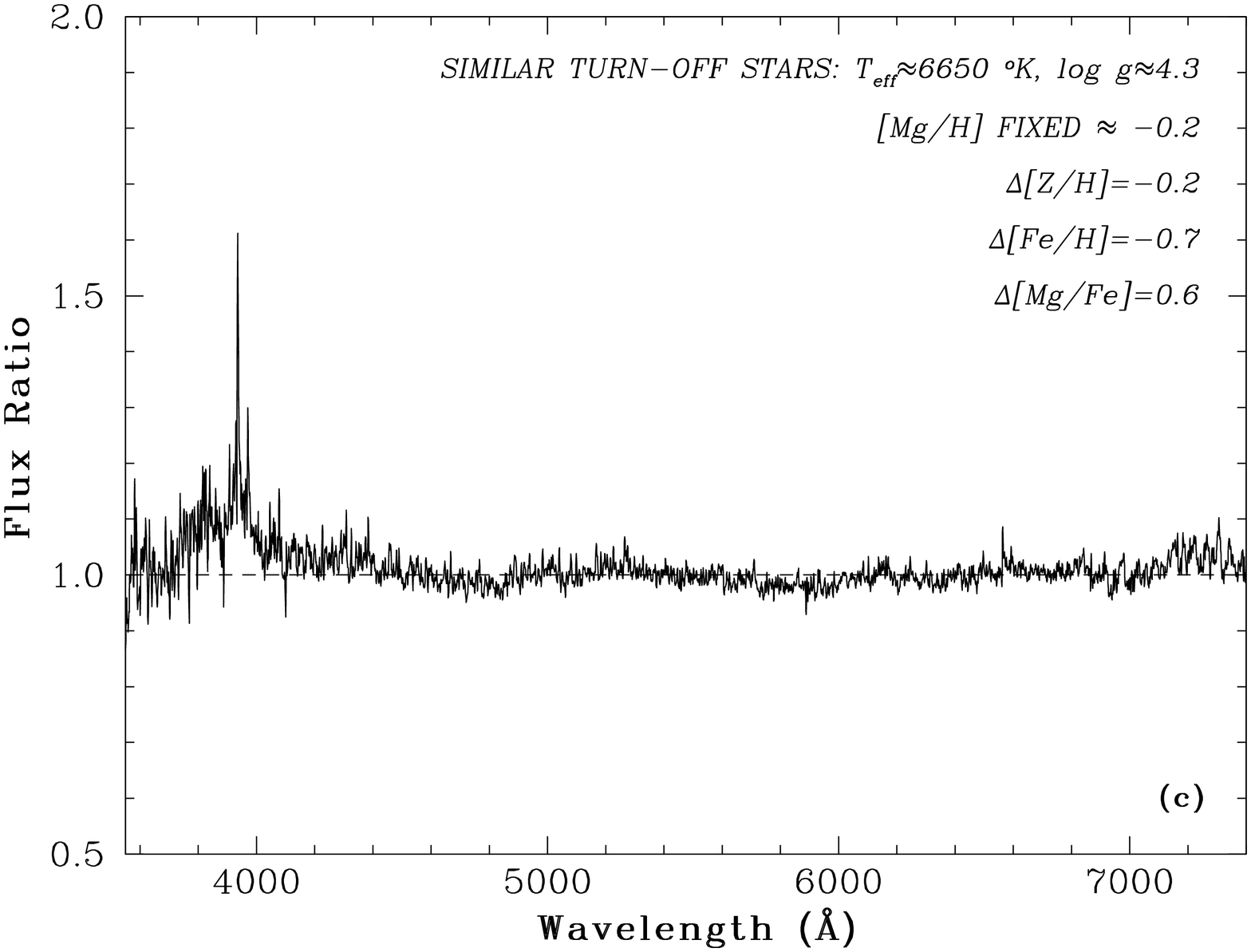}
\caption{ 
Ratios of pairs of MILES spectra  
a) of a pair of analogue stars for the TO stage, 
fixing [Z/H] below the solar value but varying [Mg/H] and [Fe/H]
($\Delta$([Mg/H]) = +0.10 and $\Delta$([Fe/H]) = -0.19 dex).
b) of a pair of analogue stars for the TO stage, 
keeping [Fe/H] fixed below the solar value but changing [Mg/H] and [Z/H]
($\Delta$([Mg/H]) = +0.42 and $\Delta$([Z/H]) = +0.44 dex).
c) of a pair of analogue stars for the TO stage, 
assuming [Mg/H] constant but varying [Fe/H] and [Z/H]
($\Delta$([Fe/H]) = -0.70 and $\Delta$([Z/H]) = -0.21 dex).
The star names and their parameters are listed in Table~5.
}
\label{FigX2} 
\end{figure}

\begin{figure}
\includegraphics[width=67mm, angle=-90]{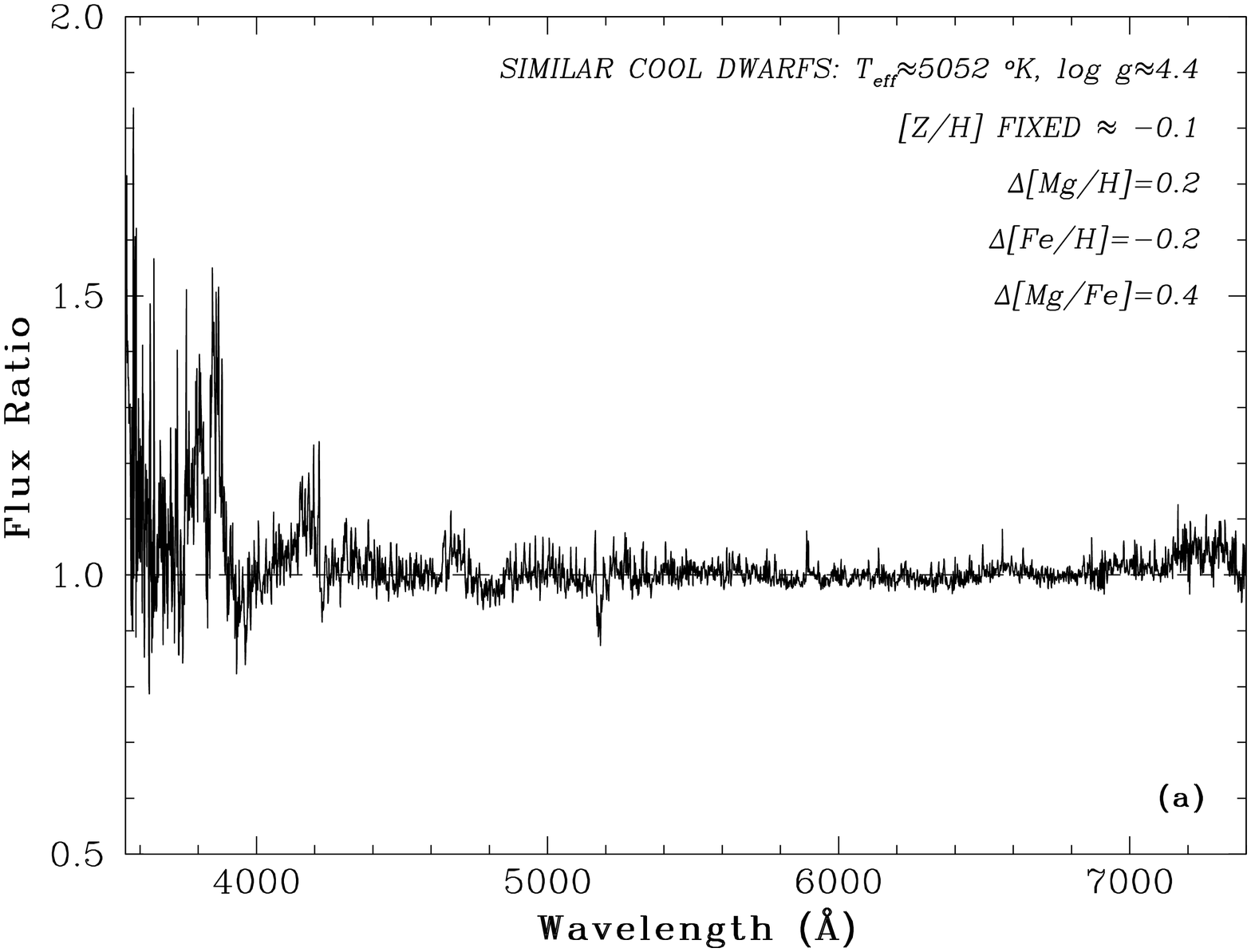}
\includegraphics[width=67mm, angle=-90]{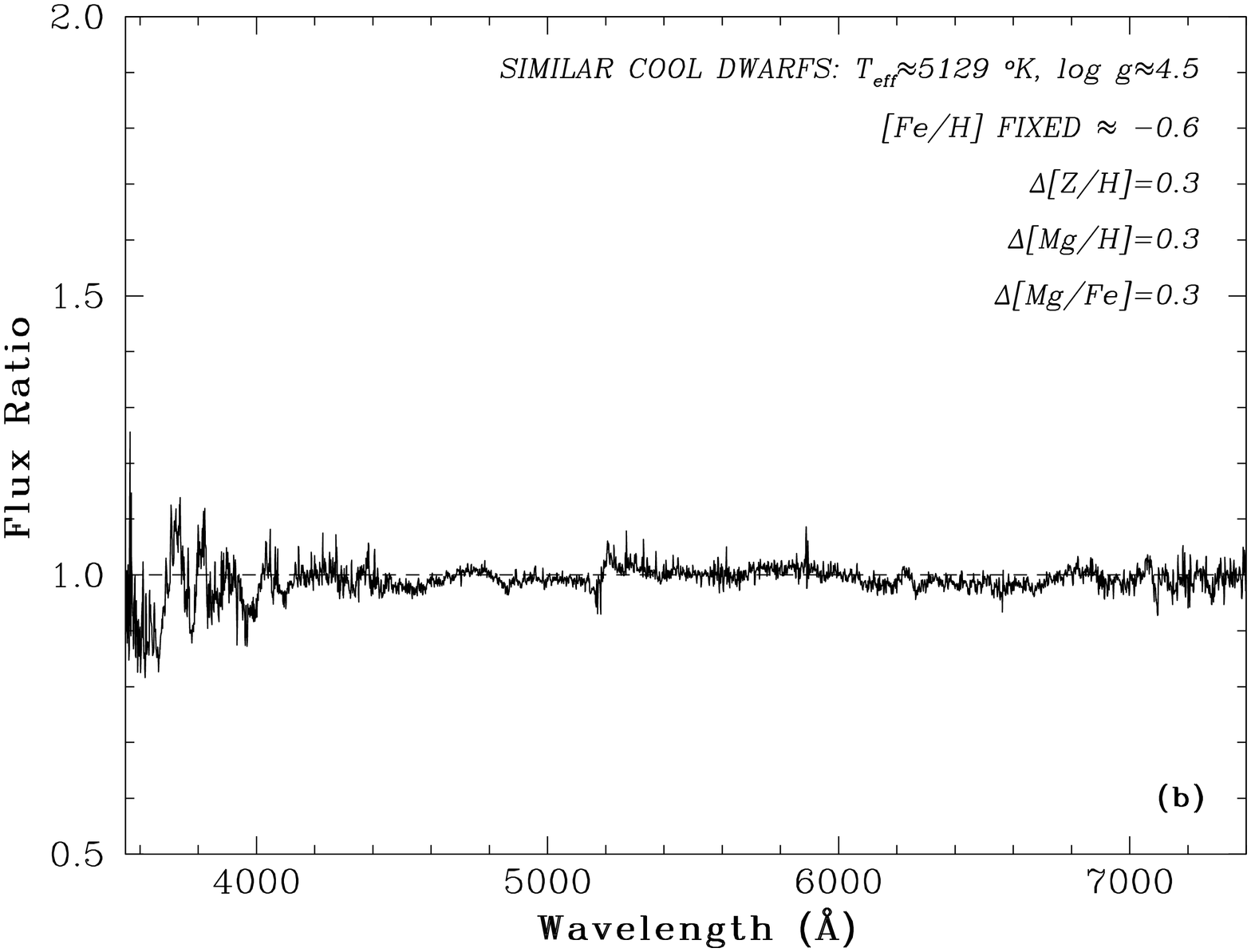} 
\includegraphics[width=67mm, angle=-90]{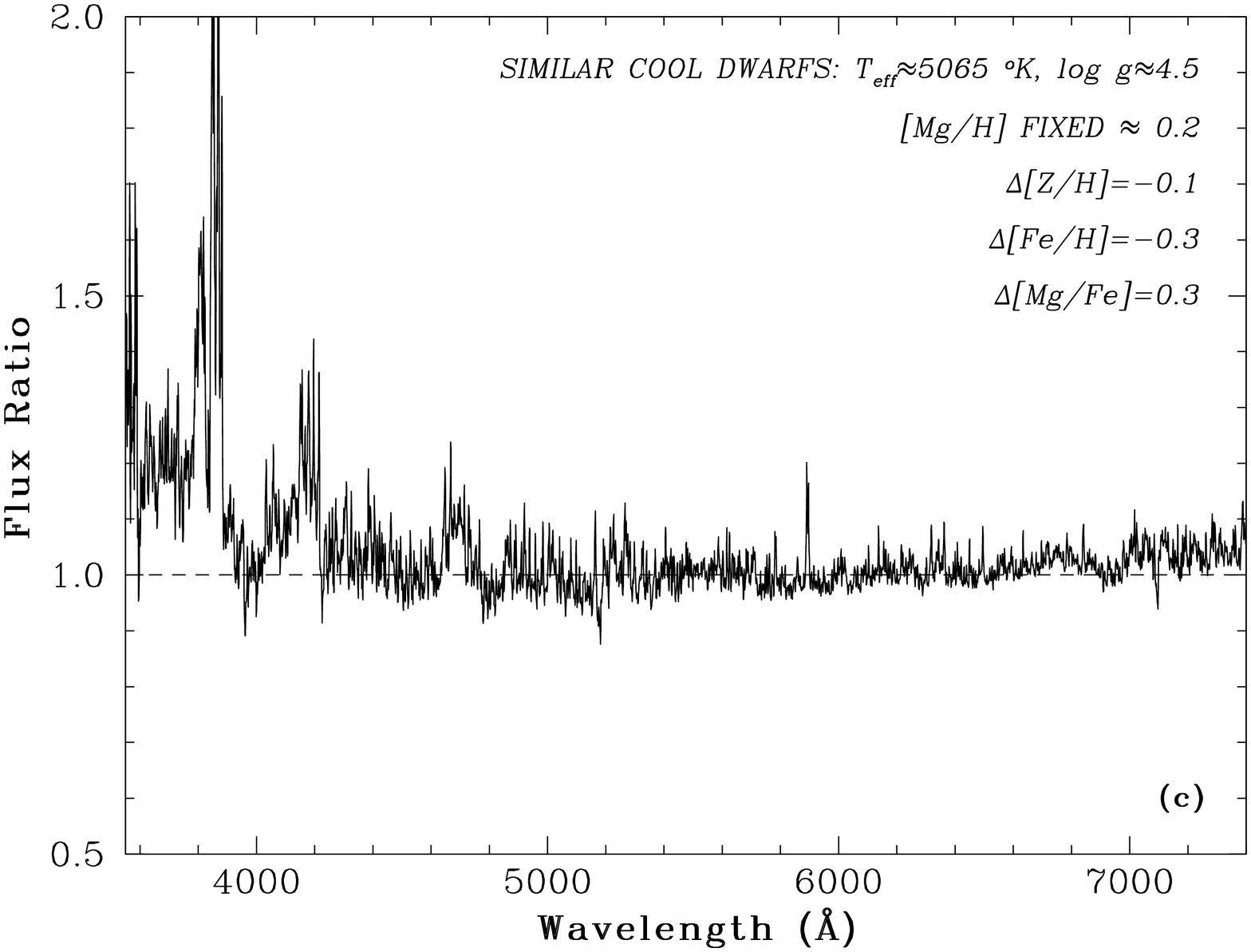}
\caption{ 
Ratios of pairs of MILES spectra  
a) of a pair of analogue stars for the CD stage, 
fixing [Z/H] around the solar value but varying [Mg/H] and [Fe/H]
($\Delta$([Mg/H]) = +0.16 and $\Delta$([Fe/H]) = -0.21 dex).
b) of a pair of analogue stars for the CD stage, 
keeping [Fe/H] fixed below the solar value but changing [Mg/Fe] and [Z/H]
($\Delta$([Mg/H]) = +0.32 and $\Delta$([Z/H]) = +0.25 dex).
c) of a pair of analogue stars for the CD stage, 
assuming [Mg/H] constant but varying [Fe/H] and [Z/H]
($\Delta$([Fe/H]) = -0.33 and $\Delta$([Z/H]) = -0.10 dex).
The star names and their parameters are listed in Table~5.
}
\label{FigX3} 
\end{figure}

\subsubsection{Effect of parameter errors on spectrum ratios}
We investigated the impact of errors in $T_{\rm eff}$ and $\log~g$ on 
flux ratios of similar stars. In the MILES database the typical uncertainty 
for FGK stars is 100K in temperature and 0.2 in $\log~g$. 
To analyse the influence of significant temperature and surface gravity
deviations we computed spectrum ratios for selected pairs of analogue stars
with very similar parameters except for temperature (which deviated 
by $\ge 275$K) or log gravity (which deviated by $\ge 0.6$).

For cool giants a temperature increase of more than 3 
times the temperature uncertainty 
produces more blue flux (from 20\% upwards) and residuals in lines 
across the spectrum, appearing all as excesses or deficiencies in the 
spectrum ratios. The difference dominates in the blue, however, 
the whole spectrum is affected to some extent.
Flux ratios analysed for pairs of RGB stars do not exhibit this pattern
due to such $T_{\rm eff}$ deviations. Also the pattern of effects 
produced by $\log~g$ uncertainties are not seen in the RGB flux ratios 
shown in Fig.~5.
For the TO stars, effects of these temperature and gravity uncertainties 
are not significant, except that Ca II H-K lines just below 4000 \AA\ 
are affected by changes in both $\log~g$ and $T_{\rm eff}$ at some level 
(perhaps affecting the spectrum ratios shown in Fig.~6, especially
Fig.~6c which shows the constant [Mg/H] case).
The impact of temperature uncertainties on the CD spectrum ratios 
exhibits qualitatively similar behaviour as in the RGB case, but with a
smaller magnitude since the CD stars are somewhat hotter. 
There is no significant effect of gravity uncertainty on flux ratios 
for the CD case, except for the constant [Fe/H] case, where 
the spectrum ratio is close to one throughout (Fig.~7b).
Therefore we are confident that the differences that we are seeing in 
Figs.~5, 6 and 7 are not dominated by $T_{\rm eff}$ and $\log~g$ 
parameter uncertainties in these MILES similar star pairs.

\subsubsection{Influence of C, N and O abundances on [Z/H] estimation}

The CNO group is an important contributor to the total metal content
and integrated opacity in a stellar phostosphere. To investigate 
the impact of individual abundances of carbon, nitrogen and oxygen 
on the global metallicity estimate, we recomputed [Z/H] on a 
star-by-star basis (for the stars listed in Table~5) adopting their 
published abundances where available. When either there is no 
elemental abundance available or the star's collected [Fe/H] does 
not match its MILES value (within 2$\sigma[Fe/H]$=0.2 dex), we 
estimated [X/Fe] from observed mean galactic trends for local disk 
stars. Table~6 compiles the individual re-estimated [Z/H] as well 
as the CNO abundances. This approach should be more precise than 
the previously applied approximation (Eq.~6), in which the 
$\alpha$-element abundances (including oxygen) are all represented 
by magnesium, with carbon and nitrogen assumed to be scaled-solar. 
The galactic trends of [C/Fe] and [N/Fe] as 
a function of [Fe/H] for dwarf stars are those from Takeda \& Honda 
(2005), i.e. [C/Fe] = -0.21($\pm$0.03)[Fe/H] + 
0.014($\pm$0.006), and [N/Fe] around the solar value (obtained 
from a sample of 160 nearby FGK dwarfs/subgiants with -0.7 $\leq$ 
[Fe/H] $\leq$ +0.4). The trends of [O/Fe] for dwarfs and [C,N,O/Fe] 
for giants are listed in Appendix B (Table~B1) and illustrated 
in Fig.~B1; they are respectively from Soubiran \& Girardi (2005) 
and Luck \& Heiter (2007).

According to our simpler procedure to estimate [Z/H], the variation 
in [Z/H] is linearly correlated to the variations in [Fe/H] and 
[$\alpha$/H] (Eq.~7). To evaluate how the individual CNO abundances 
modify this approximation, we checked if $\Delta$[Z/H], 
$\Delta$[Fe/H] and $\Delta$[$\alpha$/H] (Mg or O as a proxy)
follow this differential relationship considering typical abundance 
errors (about 0.1 dex on average). In summary our main results are 
as follows for the pairs of similar stars given in Table~5 and 
the differences in estimated [Z/H] values are plotted in Fig.~8.
\begin{itemize}
\item [[Z/H]] constant:
$\Delta$[Fe/H], accounting for the small variations in
[Z/H] (of 0.08 and 0.13 dex in the RG and TO stars respectively), 
is better correlated with $\Delta$[O/H] instead of $\Delta$[Mg/H]. 
For the CD case, the variation in [Z/H] is very close to zero.
\item [[Fe/H]] constant:
The expected correlation $d[Z/H] = 0.75d[\alpha/H]$ 
from Eq.~7 works (within the abundance 
uncertainties), except that the two RG stars show the largest 
deviations, as seen plotted in Fig.~8.
\item [[$\alpha$/H]] constant:
The variation in [Fe/H] correlates well with $\Delta$[Z/H] following 
our simple approximation in Eq.~7. For the TO case, the 
relation would be better reproduced if [$\alpha$/H] differences, 
in elements other than Mg, were allowed for. In the CD case, the 
differential relationship (Eq.~7) would be acceptable if the 
outlying data from Zhao et al. (2002) were excluded 
(see their [O/H] value in Table~6).
\end{itemize}

In general we find that the few available C, N \& O individual 
abundances have some influence on the estimation of overall 
metallicity [Z/H], but that it is not a significant affect, taking 
into account the abundance uncertainties (there are only two stars 
that deviate from Eq.~7 by 2 to 3$\sigma[Fe/H]$, both from the [Fe/H] 
constant case of RG stars, where [O/H] does not follow [Mg/H] - 
see Fig.~8). Only one stellar spectral comparison 
is probably invalid (the [$\alpha$/H] constant case of TO stars). 
Thus our standard approach expressed by Eqs.~6 \& 7 can be considered 
as a reliable approximation of [Z/H] for the present analysis. 
However adopting well-determinate CNO abundances in a homogeneous 
system will provide more precise global metallicity estimates in future.
We will be able to redo the spectral-ratio analysis when we have 
completed an abundance compilation for as many MILES stars as possible.
An important aspect of this task will be to transform all [X/Fe]
onto a uniform system, checking the scales of T$_{\rm eff}$, 
$\log$~g and [Fe/H] of each referenced work against the MILES parameters 
system. This is a longer-term project that is currently in progress.

\begin{figure}
 \centering
 \includegraphics[width=60mm, angle=-90]{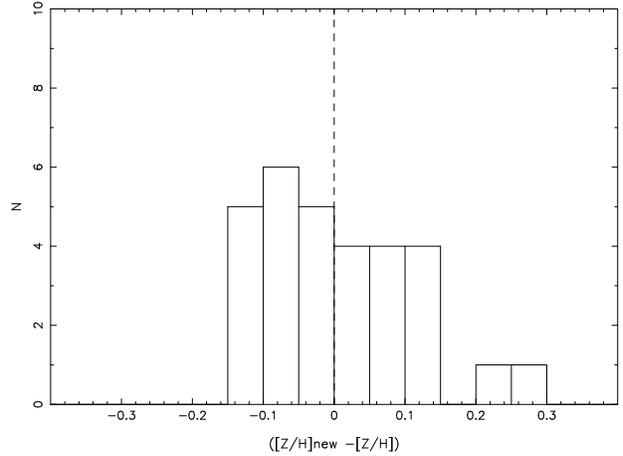} 
 \caption{Histogram of differences between [Z/H]new (derived using 
individual C, N, O, Mg and Fe abundances from Table~6) and [Z/H] estimated 
using Eq.~6.} 
\end{figure}

\begin{table*} 
 \centering
 \begin{minipage}{175mm}
 \begin{center}
\caption{C, N and O data for pairs of similar stars (corresponding to those in Table 5).}
\begin{tabular}{lrrrrrrrrrrrrrlr}
Name     &$[{{\rm Fe}\over{\rm H}}]$ &$[{{\rm Mg}\over{\rm Fe}}]$ &$[{{\rm Mg}\over{\rm H}}]$ &$[{{\rm Fe}\over{\rm H}}]$ &$[{{\rm C}\over{\rm Fe}}]$ &$[{{\rm C}\over{\rm H}}]$ &Ref   &$[{{\rm N}\over{\rm Fe}}]$ &$[{{\rm N}\over{\rm H}}]$ &Ref   &$[{{\rm O}\over{\rm Fe}}]$ &$[{{\rm O}\over{\rm H}}]$ &Ref &Notes &$[{{\rm Z}\over{\rm H}}]$ \\
         & \multicolumn{3}{c}{---------MILES---------} &       &       &      &or    &       &      &or    &       &      &or     &        & new \\
         &       &        &       &       &       &      &Trend &       &      &Trend &       &      &Trend  &      &      \\
         &  dex  &  dex   & dex   & dex   & dex   & dex  &      & dex   & dex  &      & dex   & dex  &       &      & dex  \\
\hline
 &&&&&&&&&&&&&&& \\
\multicolumn{16}{c}{=RED GIANTS=} \\
\multicolumn{16}{l}{[Z/H] constant} \\
HD192909 & -0.43 &  0.53  & 0.10  &       & 0.03  &-0.40 &Trend & 0.15  &-0.28 &Trend & 0.38  &-0.05 &Trend  &      &-0.10 \\
HD164058 & -0.05 &  0.02  &-0.03  &       &-0.14  &-0.19 &Trend & 0.20  & 0.15 &Trend & 0.07  & 0.02 &Trend  &      &-0.02 \\
 &&&&&&&&&&&&&&& \\
\multicolumn{16}{l}{[Fe/H] constant} \\
HD009138 & -0.37 &  0.19  &-0.18  &-0.34  & 0.19  &-0.15 & LH07 & 0.15  &-0.19 & LH07 & 0.52  & 0.18 & LH07  & CNO  & 0.01 \\
HD137704 & -0.37 & -0.16  &-0.53  &       & 0.01  &-0.36 &Trend & 0.16  &-0.21 &Trend & 0.32  &-0.05 &Trend  &      &-0.19 \\
 &&&&&&&&&&&&&&& \\
\multicolumn{16}{l}{[Mg/H] constant} \\
HD192909 & -0.43 &  0.53  & 0.10  &       & 0.03  &-0.40 &Trend & 0.15  &-0.28 &Trend & 0.38  &-0.05 &Trend  &      &-0.10 \\
HD139669 & -0.01 &  0.06  & 0.05  &       &-0.15  &-0.16 &Trend & 0.20  & 0.19 &Trend & 0.05  & 0.04 &Trend  &      & 0.01 \\
 &&&&&&&&&&&&&&& \\
 &&&&&&&&&&&&&&& \\
\multicolumn{16}{c}{=TO STARS=} \\
\multicolumn{16}{l}{[Z/H] constant} \\
HD109443 & -0.65 &  0.43  &-0.22  &       & 0.15  &-0.50 &Trend & 0.00  &-0.65 &Trend & 0.25  &-0.40 &Trend  &      &-0.40 \\
HD130817 & -0.46 &  0.14  &-0.32  &       & 0.11  &-0.35 &Trend & 0.00  &-0.46 &Trend & 0.26  &-0.20 &GL93   & O    &-0.27 \\
 &&&&&&&&&&&&&&& \\
\multicolumn{16}{l}{[Fe/H] constant} \\
HD119288 & -0.46 &  0.53  & 0.07  &       & 0.11  &-0.35 &Trend & 0.00  &-0.46 &Trend & 0.25  &-0.21 &Trend  &*     &-0.19 \\
HD099747 & -0.51 &  0.16  &-0.35  &       & 0.12  &-0.39 &Trend & 0.00  &-0.51 &Trend & 0.25  &-0.26 &Trend  &      &-0.32 \\
 &&&&&&&&&&&&&&& \\
\multicolumn{16}{l}{[Mg/H] constant} \\
HD109443 & -0.65 &  0.43  &-0.22  &       & 0.15  &-0.50 &Trend & 0.00  &-0.65 &Trend & 0.25  &-0.40 &Trend  &      &-0.40 \\
HD125451 &  0.05 & -0.22  &-0.17  &       & 0.00  & 0.05 &Trend & 0.00  & 0.05 &Trend &-0.19  &-0.14 &GL93   & O    &-0.07 \\
 &&&&&&&&&&&&&&& \\
 &&&&&&&&&&&&&&& \\
\multicolumn{16}{c}{=COOL DWARFS=} \\
\multicolumn{16}{l}{[Z/H] constant} \\
HD026965 & -0.31 &  0.34  & 0.03  &-0.24  & 0.14  &-0.10 &LH06  & 0.00  &-0.31 &Trend & 0.12  &-0.12 &LH06   &CO    &-0.15 \\
         &       &        &       &-0.28  & 0.08  &-0.23 &Trend & 0.00  &-0.31 &Trend & 0.38  & 0.10 &PM11   &O     &-0.02 \\
         &       &        &       &-0.31  & 0.08  &-0.23 &Trend & 0.00  &-0.31 &Trend & 0.41  & 0.10 &Re07   &O     & 0.00 \\
         &       &        &       &-0.31  & 0.42  & 0.11 &Ee04  & 0.00  &-0.31 &Trend & 0.23  &-0.08 &Trend  &C     &-0.02 \\
HD171999 & -0.10 & -0.03  &-0.13  &       & 0.03  &-0.07 &Trend & 0.00  &-0.10 &Trend & 0.16  & 0.06 &Trend  &**    &-0.01 \\
 &&&&&&&&&&&&&&& \\
\multicolumn{16}{l}{[Fe/H] constant} \\
HD132142 & -0.55 &  0.34  &-0.21  &-0.54  & 0.13  &-0.42 &Trend & 0.00  &-0.55 &Trend & 0.24  &-0.30 &Ce06   &$\alpha$  &-0.33 \\
         &       &        &       &-0.45  & 0.13  &-0.42 &Trend & 0.00  &-0.55 &Trend & 0.51  & 0.06 &PM11   &O         &-0.17 \\
         &       &        &       &       & 0.13  &-0.42 &Trend & 0.00  &-0.55 &Trend & 0.25  &-0.30 &Trend  &          &-0.33 \\
HD025673 & -0.60 &  0.07  &-0.53  &-0.53  & 0.14  &-0.46 &Trend & 0.00  &-0.60 &Trend & 0.15  &-0.38 &Ce06   &$\alpha$  &-0.48 \\
         &       &        &       &-0.50  & 0.32  &-0.18 &DM10  & 0.00  &-0.60 &Trend & 0.15  &-0.35 &DM10   &CO        &-0.42 \\
         &       &        &       &       & 0.14  &-0.46 &Trend & 0.00  &-0.60 &Trend & 0.25  &-0.35 &Trend  &          &-0.42 \\   
 &&&&&&&&&&&&&&& \\
\multicolumn{16}{l}{[Mg/H] constant} \\
HD190404 & -0.17 &  0.39  & 0.22  &       & 0.05  &-0.12 &Trend & 0.00  &-0.17 &Trend & 0.19  & 0.02 &Trend  &      & 0.02 \\
HD075732 &  0.16 &  0.09  & 0.25  & 0.32  &-0.02  & 0.14 &Trend & 0.00  & 0.16 &Trend & 0.04  & 0.36 &Ze02   &O     & 0.19 \\
         &       &        &       & 0.31  &-0.02  & 0.14 &Trend & 0.00  & 0.16 &Trend &-0.18  & 0.13 &PM11   &O     & 0.09 \\
         &       &        &       & 0.33  &-0.02  & 0.14 &Trend & 0.32  & 0.65 &Ee04  &-0.05  & 0.11 &Trend  &N     & 0.18 \\
         &       &        &       & 0.33  &-0.02  & 0.14 &Trend & 0.00  & 0.16 &Trend &-0.20  & 0.13 &Ee06   &O     & 0.09 \\
         &       &        &       &       &-0.02  & 0.14 &Trend & 0.00  & 0.16 &Trend &-0.05  & 0.11 &Trend  &      & 0.15 \\
 &&&&&&&&&&&&&&& \\
\hline
\end{tabular}
\end{center}
ADDITIONAL NOTES FOR TABLE 6.
{\it Reference:} LH07=Luck \& Heiter 2007; GL93=Garcia Lopez et al. 1993; LH06=Luck \& Heiter 2006; 
PM11=Petigura \& Marcy 2011; Re07=Ram\'\i rez et al. 2007; Ee04=Ecuvillon et al. 2004; 
Ce06=Casagrande et al. 2006; DM10=Delgado Mena et al. 2010; Ze02=Zhao et al. (2002); 
Ee06=Ecuvillon et al. (2006).

(i) For galactic trend estimates only: [X/H] = [Fe/H]$_{\rm MILES}$ + [X/Fe]$_{\rm Trend}$;
otherwise [X/H] = [Fe/H]$_{\rm Ref}$ + [X/Fe]$_{\rm Ref}$.

(ii) Mean galactic trends of [X/Fe] as a function of [Fe/H] for local disk stars:
Giants
C \& N: Takeda \& Honda (2005); O: Soubirane \& Girard (2005)
Dwarfs
C, N and O: LH07

(iii) * 1 work, [Fe/H] deviates from MILES (HD119288): Clementini et al. (1999)

(iv) ** 2 works, [Fe/H] deviates from MILES (HD171999): PM11, and Trevisan et al. (2011)

\end{minipage}
\label{star_pairs_CNOindividual}
\end{table*}

\section[]{Discussion}
Uncertainties in response functions may lead to different predictions for
stellar population ages as well as abundances. For example, for a CG star,
an increase in [$\alpha$/Fe] of +0.3 at fixed 
[Fe/H]=0.0 leads to predicted changes in H$\delta_F$ of +0.56\AA\ 
(from K05 response functions) and +0.36\AA\ (from H02 response functions),
a difference in predictions of $\Delta$H$\delta_F$=+0.20\AA.
Alternatively, using response functions for overall [Z/H] then lowering the 
Fe-peak elements and Carbon back down to solar leads to 
predicted changes in H$\delta_F$ of -0.08\AA\ (from K05 response functions) 
and +0.32\AA\ (from H02 response functions). The difference between these 
predictions is thus $\Delta$H$\delta_F$=+0.40\AA. This is significant when 
compared with changes in H$\delta_F$ expected with age in SSPs (at 5 Gyr, 
[Fe/H]=0.0), as shown in \citet{b21}, their fig.~7:
age increases by $\sim$3 Gyr for a drop of 0.4\AA\ in H$\delta_F$.
Thus the larger predicted increase in H$\delta_F$ from K05 response functions 
would result in a slightly older age estimate, since more of the H$\delta_F$ 
increase is explained away as due to abundance ratio effects in this case.

This effect is diluted when a range of stellar types is considered in the 
calculations. Following the luminosity weighting combination used 
by \citet{b282} 
(53, 44 and 3\% of the light from CG, TO and CD stars respectively, 
approximating a 5 Gyr population), we raise only the $\alpha$-element group 
by +0.3 in the log and find a difference of $\Delta$H$\delta_F$=+0.07\AA, 
between K05 and H02 response function predictions. 
This corresponds to a change in age of less than 1 Gyr. Larger differences 
between K05 and H02 predictions are found when the [Z/H] column of the 
response functions is used (as discussed in Section~3 above), 
which can lead to significant age uncertainties for an SSP.

Deviations for the higher order Balmer features in cool stars, 
seen in Fig.~1(b), correlate more strongly with the metallicity of 
the stars (characterised by [Fe/H]) than they do with [Mg/Fe]. We
used the column for overall [Z/H] changes in the response functions 
tested in Figures 1 \& 2, in order to reach the correct [Fe/H] values 
(for solar abundance ratios), before modifying the index changes 
due to non-solar abundance ratios using the $\alpha$-element columns 
of the response function tables.
Therefore, it is likely that the most uncertain response function 
predictions for these features in K05 are the ones tabulated for 
[Z/H] changes. More accurate theoretical predictions for these changes 
are needed in the blue part of the spectrum in order to make accurate 
predictions for how H$\gamma$ and H$\delta$ absorption features should 
vary with overall metallicity and with [Fe/H].

Another area of uncertainty is how individual elements may vary on a 
star-to-star basis and the effect that this may have on the current 
comparisons. To address this question more accurately, it will be 
important in future work to obtain high spectral resolution 
observations for all these tested stars.

At present the best agreement is with the H02 response functions 
for the higher order Balmer features. Therefore the use of these is 
recommended, particularly for age determinations using these features. 
In future, more comprehensive response functions are needed for a 
wider range of star types, utilising more accurate theoretical 
predictions in the blue part of the spectrum.

H$\beta$ shows larger than expected scatter, particularly for the TO stars 
(green triangles in Fig.~1(b)). The sensitivity of this index to abundance 
pattern variations needs further study, since conflicting results exist 
between current theoretical models \citep[e.g.][]{b501,b151}. 

Another important feature, whose behaviour with abundance pattern 
variations is not well reproduced by the response functions of K05 
(or H02) is Ca4227, which is sensitive to calcium. There is a large 
scatter between theoretical predictions and empirical measurements 
for this feature (see Fig.~1(e)). This feature has been used in the 
past to conclude that giant ellipticals are under-abundant in calcium 
(i.e [Ca/Mg]$<$0.0) and hence that calcium follows iron more closely in 
those galaxies \citep{b302,b2002}. 
However, the lack of good predictions of Ca4227 line 
strengths in stars in the local solar neighbourhood, as seen in Fig.~1(e), 
calls into question the accuracy of the response functions for this 
feature. This feature is thought to be affected by CN bands \citep{b181}.
Therefore, for its accurate interpretation, it may be that the CN band 
strength also needs to be accurately predicted, and any assumption about 
the behaviour of C or N may lead to inaccurate conclusions about the 
interpretation of the Ca4227 line strength. There is a weak trend of
increasing offsets below the 1:1 line, with increasing [Mg/Fe] for Ca4227, 
which also hints at additional abundance dependencies that are not yet 
fully accounted for in the response functions for this feature.
The magnesium sensitive 
features (Mg$_2$ and Mgb) show more of a correlation with theoretical 
expectations (in Fig.~1(e) and Fig.~2 for K05 and H02 response 
functions respectively). However, there is still some residual scatter, 
which is unexplained by the abundance patterns assumed here and may 
point to more complex abundance pattern variations between stars.

The differences in response functions for the higher order Balmer features,
from different theoretical models, lead to uncertainties in both ages 
and chemistry of stars and stellar populations. This is an additional 
uncertainty not normally taken into account in papers that publish stellar 
population parameters and draw conclusions from Lick indices fitting. 
As indicated earlier, there is a move towards generation of whole spectral 
SSPs and fitting of such to data, rather than using indices. This full 
spectrum fitting approach will also be affected by any mismatches between 
theoretical predictions and empirical observations. It is recommended that 
future generations of SSP model producers, of indices or spectra, test 
their results on a star-by-star basis against observations for a 
range of star abundance patterns (i.e. a range of [Fe/H] and [$\alpha$/Fe]), 
in order to check for any discrepancies in the predictions, like 
those found here for the higher order Balmer features and for other features
(e.g. Ca4227). It is particularly important to check against empirical 
measurements of indices, since those isolate the parts of spectra that help 
most to break the well known degeneracies, and to isolate features most 
sensitive to particular element abundances.

\begin{figure}
\includegraphics[width=67mm, angle=-90]{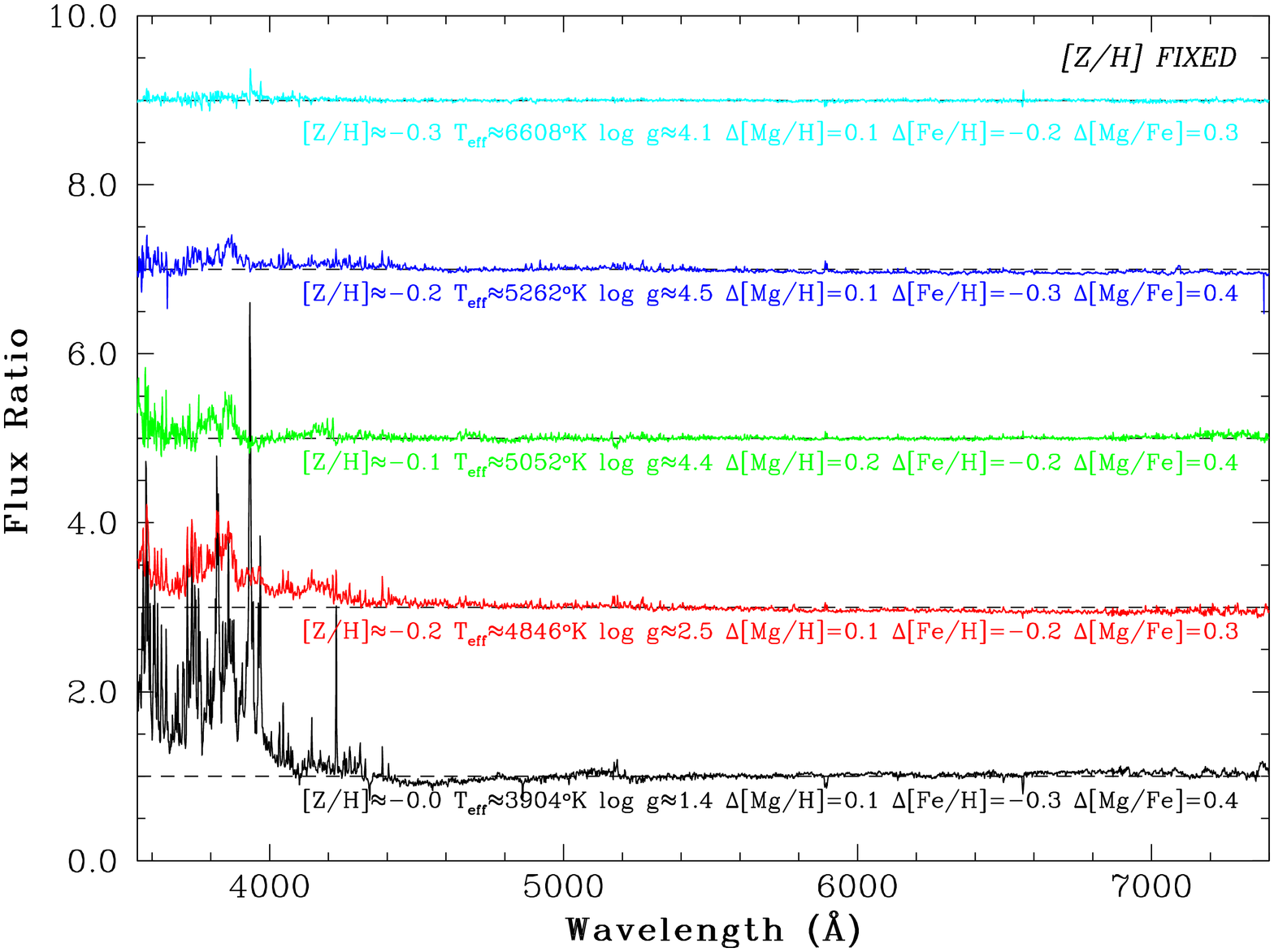}
\caption{ 
Ratios of pairs of MILES spectra showing the increasing importance 
of abundance pattern variations for cooler temperature stars.
The plots are at fixed [Z/H] (typically around -0.2, estimated
using equation 6), but varying [Mg/H] and [Fe/H] and are successively 
offset by 2.0 to avoid overlap. The star parameters are listed under each plot.
}
\label{FigT} 
\end{figure}

The spectral ratios shown in Section 4 illustrate that the impact of 
abundance variations on the blue region of the spectrum decreases with 
increasing temperature. This is seen when ratios of stars at fixed 
[Z/H], but varying [Fe/H] and [Mg/H] are plotted in order of increasing 
temperature in Fig.~9, which show decreasing variations with increasing 
temperature. This result is larger than the uncertainties due to 
stellar parameters, assessed from studies of similar stars.
Therefore, study of abundance effects in the blue region of the 
spectrum is particularly important for cool stars.

\section[]{Conclusions}
  The effects of element abundance changes on the strengths of spectral 
features in the spectra of different types of stars have been investigated.
Theoretical response functions, widely used to measure abundance 
patterns in observed stellar populations, are tested against empirical
data for stars from the MILES stellar library with measured abundances of 
[Fe/H] and [Mg/Fe]. Using the empirical [Mg/Fe] measurements from M11 
as a proxy for overall [$\alpha$/Fe] values, the following results are 
found from these tests.
\begin{enumerate}
  \item For K05 response functions the Fe sensitive features largely 
follow the observations, whereas H$\gamma$ and H$\delta$ features show 
systematically different behaviour between theoretical predictions and 
empirical observations. For the H$\gamma$ and H$\delta$ features, warm stars 
show a wider range of indices than predicted, whereas the opposite is 
true for cool stars. Indices sensitive to other elements show weaker 
trends, with larger scatter about the one-to-one lines (e.g. Mgb and Mg$_2$). 
The calcium sensitive feature (Ca4227) shows negligible trend, implying 
that additional factors affect this index apart from overall [$\alpha$/Fe].
  \item For H02 response functions similar results are found as for the K05 
comparisons, however, the agreement between theory and observations is 
improved for the H$\gamma$ and H$\delta$ features in cool stars 
when H02 response functions are used.
  \item It is important to compile and obtain results for [Ca/Fe] 
measurement 
for MILES stars in future, to gain a better understanding of the influence 
of calcium on specific features such as Ca4227. Future measurements of 
carbon and nitrogen abundances in the stars studied in this paper are also 
important to obtain from high resolution spectra, in order to better 
understand the element responses of CN$_1$, 
CN$_2$, Ca4227 and other features.
  \item For W12 star response functions (used in L09), together with K05 
[Z/H] overall metallicity responses, similar patterns are found. These W12 
star response functions could also be used for comparing many more stars 
to explore small changes in [$\alpha$/Fe] only. This showed typically a 
large scatter between normalised observations and normalised theoretical 
predictions, with weak trends about the one-to-one line for 
Mg$_2$, Mgb and NaD indices. 
  \item Full spectrum comparisons show that changes in the blue part of 
the spectrum are largely due to changes in [Fe/H] abundance. These changes 
decrease with increasing star temperature. 
  \item Overall [Z/H] is not always the most appropriate way to 
rank stars, since abundances of individual elements have important effects 
on emergent spectra and in general this is particularly important in the 
blue part of the spectra of cool stars, where Fe, C and N abundances 
strongly affect the spectral shape.
  \item The spectral results so far indicate the need for deeper observational 
and theoretical studies of the 
blue part of stellar spectra, to search for more measurable metallicity 
indicators, senitive to iron abundance and to other element abundances in 
different types of stars. For example, indices in the blue have been defined 
by \citet{b19,b22}. 
We will explore this direction in a future paper.
\end{enumerate}

  In summary, this current work shows that theoretical response functions 
of K05 and H02 work quite
well for most Lick spectral indices, with the exception of systematic 
offsets in the H$\gamma$ and H$\delta$ features, when compared to observed 
stars. This effect is important for individual stars and to a lesser 
extent for stellar population analysis, where the opposite systematics of 
warm and cool stars partially compensate for each other.
The response functions need to be applied in a careful 
and limited way, taking into account the expected spread of values and 
types of indices, on an index-by-index basis. If response functions are 
applied automatically in a single method and for unlimited abundance 
variations, then they will produce spurious results in derived abundance 
patterns and in stellar population ages. 

\section*{Acknowledgments}

We thank Paula Coelho for insightful discussions and comments on this manuscipt.
We thank the IAC and UCLan for travel funding facilitating the work for 
this paper. A Brazilian grant from CAPES (Coordena\c c\~ao de Aperfei\c 
coamento de Pessoal de N\'\i vel Superior) was awarded for AdCM to visit UCLan.
We also thank the Brazilian funding organisation FAPESP 
(Funda\c c\~ao de Amparo \`a Pesquisa do Estado de S\~ao Paulo)
for travel funding support for AES (grant number 2012/04953-0).
This work has been supported by the Programa Nacional de 
Astronom{\'{\i}}a y Astrof{\'{\i}}sica of the Spanish Ministry 
of Economy and Competitiveness (MINECO) un-der grant AYA2010-21322-C03-02.
Thanks to G. Worthey for providing us with information and values for 
their star response functions and to B. Barbuy for software to compute overall 
abundances from an abundance pattern. Finally, we thank the anonymous referee,
who pointed out additional aspects for consideration in presenting this work.

\appendix

\section[]{Parameters for three categories of MILES stars.}

This appendix shows tables of data for 7 CD, 31 TO and 13 CG stars, 
used in testing the K05 and H02 response functions in this paper,
and plotted in Figs.~1 \& 2. Details of these measurements are given
Section 2.3. The [Mg/Fe] ratios are from M11. All 25 Lick indices are 
available in the on-line version.

\begin{table*} 
 \centering
 \begin{minipage}{175mm}
 \begin{center}
\caption{Tables of data for stars corresponding to the base model $T_{\rm eff}$ and $\log~g$ values in K05, for CD, TO and CG stars, from a 5Gyr old population.'M' in column 7 denotes the MILES star number. All 25 Lick indices are available in the on-line version.}
\begin{tabular}{llllrrlcccccccc}

\multicolumn{15}{l}{CD stars. Model=($T_{\rm eff}$=4575.0,$\log~g$=4.60)} \\
\hline
    No.& Name         & $T_{\rm eff}$& $\log~g$& [Fe/H]& [Mg/Fe]&      M&    H$\delta_A$&    H$\delta_F$&     CN$_1$&     CN$_2$&  Ca4227&   G4300&    H$\gamma_A$&    H$\gamma_F$ \\
\hline
  1& HD032147     &4658& 4.47&  0.020& -0.056& 168& -7.127& -1.370&  0.101&  0.150&  3.527&  5.809&-10.765& -3.638 \\
  2& HD131977     &4501& 4.70&  0.020&  0.124& 532& -6.714& -1.176&  0.025&  0.076&  4.021&  5.471&-10.360& -3.266 \\
  3& HD156026     &4541& 4.54& -0.370&  0.157& 625& -5.710& -0.976& -0.021&  0.034&  4.642&  4.967& -9.665& -3.166 \\
  4& HD103932     &4510& 4.57&  0.160& -0.049& 426& -6.926& -1.156&  0.058&  0.113&  4.356&  5.620&-11.064& -3.831 \\
  5& BD+430699    &4608& 4.52& -0.600&  0.237& 115& -5.430& -1.000& -0.002&  0.038&  3.183&  5.331& -8.960& -3.053 \\
  6& HD021197     &4616& 4.59&  0.300& -0.098& 117& -6.463& -1.072&  0.049&  0.103&  4.491&  5.558&-11.053& -3.551 \\
  7& HD108564     &4594& 4.67& -1.090&  0.516& 442& -3.606& -0.158& -0.016&  0.026&  3.105&  4.949& -7.847& -2.786 \\
\hline
\multicolumn{15}{l}{TO stars. Model=($T_{\rm eff}$=6200.0,$\log~g$=4.10)} \\
\hline
    No.& Name         & $T_{\rm eff}$& $\log~g$& [Fe/H]& [Mg/Fe]&      M&    H$\delta_A$&    H$\delta_F$&     CN$_1$&     CN$_2$&  Ca4227&   G4300&    H$\gamma_A$&    H$\gamma_F$ \\
\hline
  1& BD+342476    &6205& 4.12& -2.050&  0.187& 491&  4.391&  3.402& -0.089& -0.057&  0.110& -0.307&  3.953&  3.419 \\
  2& HD000400     &6205& 4.12& -0.330&  0.106&   7&  2.798&  2.492& -0.083& -0.057&  0.480&  2.991&  0.674&  2.028 \\
  3& HD009826     &6134& 4.09&  0.110&  0.115&  63&  2.318&  2.270& -0.080& -0.053&  0.650&  3.500& -0.108&  1.831 \\
  4& HD014938     &6153& 4.04& -0.350&  0.115&  86&  2.573&  2.427& -0.071& -0.042&  0.393&  2.536&  0.672&  2.124 \\
  5& HD016673     &6253& 4.28&  0.050&  0.045&  92&  2.721&  2.487& -0.085& -0.055&  0.597&  3.067&  0.614&  2.076 \\
  6& HD043318     &6224& 3.93& -0.150&  0.059& 213&  2.861&  2.550& -0.081& -0.054&  0.337&  2.866&  1.055&  2.356 \\
  7& HD074000     &6166& 4.19& -2.020&  0.377& 310&  3.832&  3.165& -0.079& -0.050&  0.160& -0.049&  3.465&  3.117 \\
  8& HD076910     &6275& 4.10& -0.500&  0.184& 328&  4.087&  3.201& -0.089& -0.056&  0.292&  1.576&  2.806&  3.126 \\
  9& HD084937     &6228& 4.01& -2.170&  0.440& 363&  4.460&  3.589& -0.091& -0.062&  0.049& -0.351&  4.089&  3.512 \\
 10& HD089744     &6219& 3.95&  0.230&  0.009& 384&  2.592&  2.361& -0.077& -0.049&  0.628&  3.356&  0.207&  2.071 \\
 11& HD097916     &6238& 4.03& -0.990&  0.454& 405&  4.690&  3.559& -0.093& -0.058&  0.264&  0.847&  3.693&  3.587 \\
 12& HD102870     &6109& 4.20&  0.170& -0.007& 422&  1.924&  2.074& -0.070& -0.043&  0.678&  3.793& -0.686&  1.556 \\
 13& HD107213     &6298& 4.01&  0.290&  0.133& 438&  2.816&  2.505& -0.082& -0.053&  0.488&  3.476&  0.321&  2.122 \\
 14& HD114642     &6249& 3.90& -0.180&  0.080& 464&  3.993&  3.011& -0.095& -0.063&  0.433&  2.310&  2.170&  2.999 \\
 15& HD142860     &6272& 4.17& -0.160&  0.070& 576&  3.045&  2.622& -0.082& -0.050&  0.477&  2.619&  1.515&  2.526 \\
 16& HD159307     &6198& 3.90& -0.730&  0.178& 635&  4.082&  3.125& -0.096& -0.063&  0.343&  1.819&  2.631&  3.000 \\
 17& HD173667     &6280& 3.97&  0.050&  0.046& 695&  3.944&  3.025& -0.097& -0.062&  0.449&  2.253&  2.471&  3.110 \\
 18& HD181096     &6276& 4.09& -0.260&  0.119& 716&  3.523&  2.879& -0.087& -0.057&  0.396&  2.400&  1.700&  2.725 \\
 19& HD215648     &6167& 4.04& -0.320&  0.172& 843&  2.769&  2.396& -0.078& -0.046&  0.459&  3.034&  0.780&  2.137 \\
 20& HD219623     &6155& 4.17& -0.040&  0.026& 868&  2.057&  2.211& -0.076& -0.049&  0.646&  3.664& -0.523&  1.641 \\
 21& HD222368     &6170& 4.09& -0.150&  0.131& 888&  2.671&  2.371& -0.083& -0.054&  0.519&  3.088&  0.616&  2.160 \\
 22& HD338529     &6165& 4.06& -2.250&  0.253& 725&  4.544&  3.532& -0.093& -0.056&  0.096& -0.399&  4.295&  3.510 \\
 23& HD097855     &6260& 4.05& -1.030&  0.003& 406&  3.625&  3.009& -0.081& -0.050&  0.391&  2.007&  2.032&  2.812 \\
 24& HD014221     &6295& 3.91& -0.350&  0.041&  83&  4.154&  3.265& -0.082& -0.050&  0.353&  1.615&  2.761&  3.366 \\
 25& BD+092190    &6270& 4.11& -2.860&  0.477& 348&  4.914&  3.817& -0.103& -0.068&  0.061& -0.825&  4.827&  3.853 \\
 26& HD089995     &6233& 3.95& -0.340& -0.046& 385&  3.907&  3.063& -0.088& -0.056&  0.306&  1.741&  2.485&  3.064 \\
 27& HD128429     &6266& 4.12& -0.130&  0.267& 518&  3.408&  2.864& -0.089& -0.054&  0.417&  2.550&  2.023&  2.834 \\
 28& HD173093     &6268& 4.09& -0.180&  0.123& 692&  3.504&  2.872& -0.086& -0.054&  0.499&  2.515&  1.961&  2.820 \\
 29& HD209369     &6288& 3.90& -0.280&  0.153& 822&  3.843&  2.851& -0.090& -0.059&  0.383&  1.938&  2.582&  3.183 \\
 30& HD218804     &6261& 4.05& -0.230& -0.054& 862&  4.078&  3.219& -0.093& -0.060&  0.464&  1.714&  2.982&  3.294 \\
 31& BD+592723    &6112& 4.17& -2.020&  0.528& 876&  3.609&  3.108& -0.073& -0.048&  0.152&  0.044&  3.090&  2.808 \\
\hline
\multicolumn{15}{l}{CG stars. Model=($T_{\rm eff}$=4255.0,$\log~g$=1.90)} \\
\hline
    No.& Name         & $T_{\rm eff}$& $\log~g$& [Fe/H]& [Mg/Fe]&      M&    H$\delta_A$&    H$\delta_F$&     CN$_1$&     CN$_2$&  Ca4227&   G4300&    H$\gamma_A$&    H$\gamma_F$ \\
\hline
  1& HD131430     &4190& 1.95&  0.100& -0.398& 528& -7.221& -1.969&  0.291&  0.341&  2.315&  6.386&-11.045& -3.348 \\
  2& HD075691     &4270& 2.12& -0.050&  0.015& 321& -6.601& -1.619&  0.220&  0.267&  1.929&  6.321& -9.967& -3.006 \\
  3& HD113092     &4283& 1.95& -0.370&  0.182& 457& -4.085& -0.698&  0.131&  0.178&  1.209&  7.001& -8.716& -2.690 \\
  4& HD191046     &4317& 2.01& -0.650&  0.474& 755& -3.513& -0.834&  0.076&  0.111&  1.036&  6.952& -7.594& -2.424 \\
  5& HD020893     &4340& 2.04&  0.080& -0.102& 114& -5.997& -1.414&  0.248&  0.295&  1.889&  6.310&-10.137& -3.109 \\
  6& HD066141     &4258& 1.90& -0.300&  0.061& 289& -5.395& -1.068&  0.167&  0.216&  1.614&  6.564& -9.383& -3.176 \\
  7& HD083618     &4231& 1.74& -0.080& -0.086& 357& -6.032& -1.450&  0.189&  0.236&  2.143&  6.322&-10.357& -3.131 \\
  8& HD124186     &4347& 2.10&  0.240&  0.002& 499& -7.452& -1.896&  0.327&  0.375&  2.077&  6.382&-10.773& -3.453 \\
  9& HD130705     &4336& 2.10&  0.410& -0.029& 526& -7.541& -1.875&  0.373&  0.423&  1.968&  6.467&-10.970& -3.532 \\
 10& HD136726     &4159& 1.91&  0.130& -0.087& 549& -6.397& -1.562&  0.217&  0.270&  2.677&  6.222&-10.563& -3.199 \\
 11& HD154733     &4200& 2.09&  0.000& -0.030& 620& -6.444& -1.494&  0.237&  0.289&  2.382&  6.048&-10.153& -3.235 \\
 12& HD171443     &4189& 1.84& -0.080&  0.009& 682& -6.517& -1.529&  0.263&  0.319&  2.168&  6.618&-10.300& -3.348 \\
 13& M67$\_$F-108    &4255& 1.84& -0.090&  0.016& 919& -6.772& -1.755&  0.238&  0.288&  2.422&  6.308&-10.212& -3.253 \\
\hline
\end{tabular}
\end{center}
\end{minipage}
\label{appendix_teststars}
\end{table*}

\section[]{Tests with different abundance ratio trends.}

For the results obtained in the main text we made the assumptions that 
all $\alpha$-element to iron ratios [$\alpha$/Fe] track the value 
of [Mg/Fe] and that carbon and nitrogen track iron. In this appendix 
we test these approximations using published data for samples of dwarf 
and giant stars. We fit mean trends to these data, to work out how 
element X varies with Mg or Fe, as a function of [Fe/H]. Table~B1 shows 
these fits, ranges and references. Fig.~B1 shows these mean relations 
in graphical form. The data from which these relations were obtained 
are plotted in the published papers (Soubiran \& Girard 2005 - hereafter 
SG05; Luck \& Heiter 2006 - hereafter LH06; Luck \& Heiter 2007 - 
hereafter LH07). These consist of at least 415 dwarf stars from SG05 
(see their table~3), 216 dwarf star from LH06 (see their tables 2 and 3), 
and $\sim$298 giant stars from LH07 (see their tables 4, 5 and 7).

\begin{table*} 
 \centering
 \begin{minipage}{175mm}
 \begin{center}
\caption{Polynomial fits for elements X showing how [X/Fe] varies 
with [Fe/H] on average, for elements modelled in the response 
function tables tested. The fitted datasets are indicated in the 
column headed 'References', the numbers of stars fitted 
are shown in the column headed '\#' and the rms deviations from 
the fit are shown in the final column, in dex.}
\begin{tabular}{lcccc}
\hline
    Polynomial fit  & [Fe/H] range  & References &  \# & rms \\
\hline
\multicolumn{4}{l}{DWARFS} \\
$[MgFe]=$ 0.0624+0.0110[Fe/H]+0.4672[Fe/H]$^2$-0.2692[Fe/H]$^3$-0.3746[Fe/H]$^4$  & -1.2 to +0.5 & SG05+LH06 & 818 & 0.096 \\
$[Ca/Fe]=$ 0.0254-0.1261[Fe/H]+0.0930[Fe/H]$^2$-0.0024[Fe/H]$^3$  & -1.2 to +0.5 & SG05+LH06 & 743 & 0.061 \\
$[Si/Fe]=$ 0.0419-0.1240[Fe/H]+0.1409[Fe/H]$^2$+0.0073[Fe/H]$^3$  & -1.2 to +0.5 & SG05+LH06 & 842 & 0.068 \\
$[Ti/Fe]=$ 0.0412-0.0649[Fe/H]+0.3384[Fe/H]$^2$+0.2060[Fe/H]$^3$  & -1.2 to +0.5 & SG05+LH06 & 731 & 0.088 \\
$[Na/Fe]=$ 0.0233+0.0757[Fe/H]+0.5329[Fe/H]$^2$+0.4258[Fe/H]$^3$  & -1.2 to +0.3 & SG05 & 567 & 0.080 \\
$[O/Fe]=$ 0.1004-0.7273[Fe/H]-1.1294[Fe/H]$^2$-0.5616[Fe/H]$^3$  & -1.2 to +0.3 & SG05 & 415 & 0.125 \\
\multicolumn{4}{l}{GIANTS} \\
$[Mg/Fe]=$ 0.0859+0.0361[Fe/H]+0.8155[Fe/H]$^2$+0.0894[Fe/H]$^3$   & -0.6 to +0.35 & LH07 & 298 & 0.103 \\
$[Ca/Fe]=$ -0.0533-0.2468[Fe/H]-0.3619[Fe/H]$^2$-0.7604[Fe/H]$^3$  & -0.6 to +0.35 & LH07 & 294 & 0.080 \\
$[Si/Fe]=$ 0.1365+0.0557[Fe/H]+0.6138[Fe/H]$^2$-0.2128[Fe/H]$^3$-0.6167[Fe/H]$^4$  & -0.6 to +0.35 & LH07 & 291 & 0.064 \\
$[Na/Fe]=$ 0.1283+0.1647[Fe/H]+0.4075[Fe/H]$^2$+0.0971[Fe/H]$^3$  & -0.6 to +0.35 & LH07 & 298 & 0.076 \\
$[O/Fe]=$ 0.0477-0.4119[Fe/H]+1.1975[Fe/H]$^2$+0.8401[Fe/H]$^3$  & -0.6 to +0.35 & LH07 & 298 & 0.110 \\
$[N/Fe]=$ 0.2060+0.1476[Fe/H]+0.0026[Fe/H]$^2$-0.1319[Fe/H]$^3$  & -0.6 to +0.35 & LH07 & 298 & 0.100 \\
$[C/Fe]=$ -0.1568-0.2862[Fe/H]+1.0187[Fe/H]$^2$+2.2720[Fe/H]$^3$+1.7822[Fe/H]$^4$  & -0.6 to +0.35 & LH07 & 298 & 0.103 \\
\hline
\end{tabular}
\end{center}
\end{minipage}
\label{appendix_meantrends}
\end{table*}

The enhancements of $\alpha$ elements (O, Ca, Si and Ti) assumed 
in the main text are modified by trends for [X/Mg], derived from 
combining: [X/Mg]=[X/Fe]-[Mg/Fe], from Table~1B. Sodium is similarly 
modified by [Na/Mg]=[Na/Fe]-[Mg/Fe]. Nitrogen is treated as enhanced 
for giants, in this appendix, and scaled-solar for dwarfs
\citep[e.g.][]{b231}.
Carbon, on the other hand, was 
originally assumed to follow iron in Section (3.1.1), which may be a 
good approximation for dwarfs (e.g. Takeda \& Honda 2005, 
Da Silva et al. 2011) but not 
for giants (LH07). Therefore we modify our assumed carbon abundances 
for giants by adding [C/Fe] (from Table~B1) to our original assumption.

\begin{figure*}
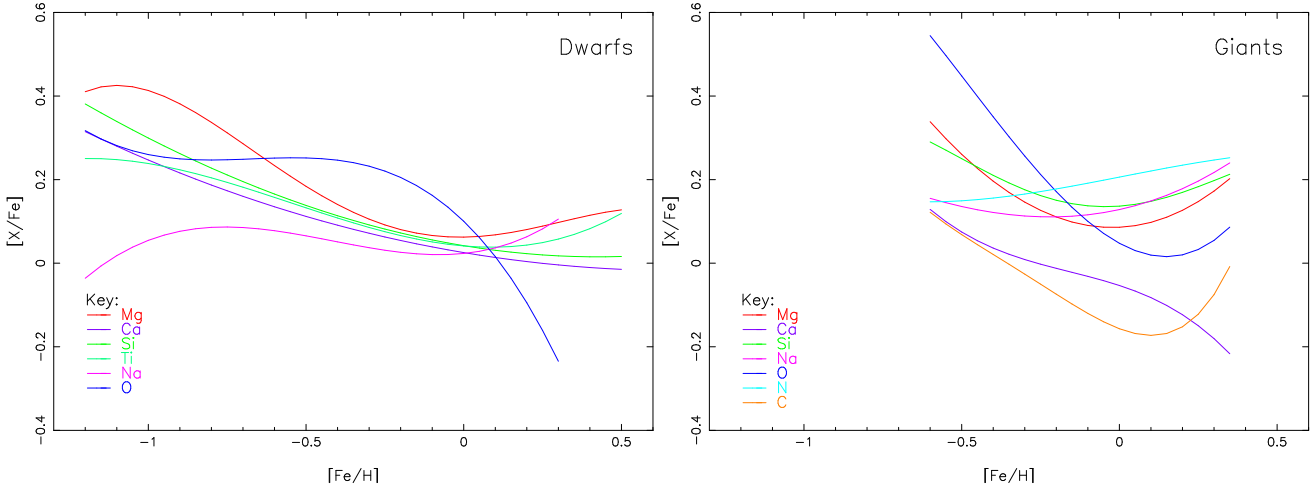

\vspace{1.0cm}
 \includegraphics[width=64mm, angle=-90]{pgplot_mean_TRENDS_Dwarfs.eps}
 \includegraphics[width=64mm, angle=-90]{pgplot_mean_TRENDS_Giants.eps}
 \caption{Mean polynomial fits for element abundance 
ratios [X/Fe] in dwarf stars (left plot) and giants stars (right plot) 
locally in the disk of the Milky Way. The horizontal ranges plotted 
illustrate ranges covered by the data (SG05+LH06 for dwarfs and LH07 
for giants).}
\end{figure*}

In this way our measured [Mg/Fe] or [Fe/H] values for each star are 
then scaled by the above observed mean trends to generate estimates 
for other elements [X/Fe]. Applying these modified abundance patterns 
leads to similar relations as seen in Figs.~1 and 2, with the main 
exceptions being cool stars in the CN$_1$ and CN$_2$ bands, Ca4227 
(affected by CN bands on one side), G4300 and to a lesser extent 
C$_2$4668. Tests of response functions for these features are therefore 
less certain, due to the greater impact of unknown C and N abundances. 
More robust tests of the responses for these 5 indices, in the blue 
part of the spectrum for cool stars, must await individual C and N 
element abundance measurements in those stars. The main results, 
regarding responses for iron features, Balmer features, magnesium 
and sodium features, remain intact. Contrast Fig.~B2, which shows 
the H$\gamma$ and H$\delta$ indices and the 4 most uncertain features, 
with the same indices plotted in Fig.~1b and in Figs.~1d,e respectively.

For H$\gamma$ and H$\delta$ indices the robustness of our findings against
uncertainties in individual element abundances points to the overall
metallicity response as the cause of the observed difference between models 
and observations in Fig.~1(b). Here we test this.
The good agreement of iron sensitive indices and others such as Ca4455
indicates that the spectral responses to overall metallicity ([Z/H]) 
in K05 and H02 are not significantly in error for those indices, in 
contrast to the case for H$\gamma$ and H$\delta$ indices. 
Table~B2 compares the overall metallicity responses for these four 
Balmer indices, from K05 and H02. From this table we see that the 
spectral responses are smaller in H02 than in K05 (except for H$\delta_A$). 
For H$\delta_A$ the overall metallicity response is larger 
in H02 and this gives a worse fit to the observations (see Table~4). 
For the other three Balmer indices in Table~B2, the smaller values of 
overall metallicity response in H02 lead to an improvement in the 
predictions for H$\gamma_A$, H$\gamma_F$ and H$\delta_F$ indices.

\begin{figure*}
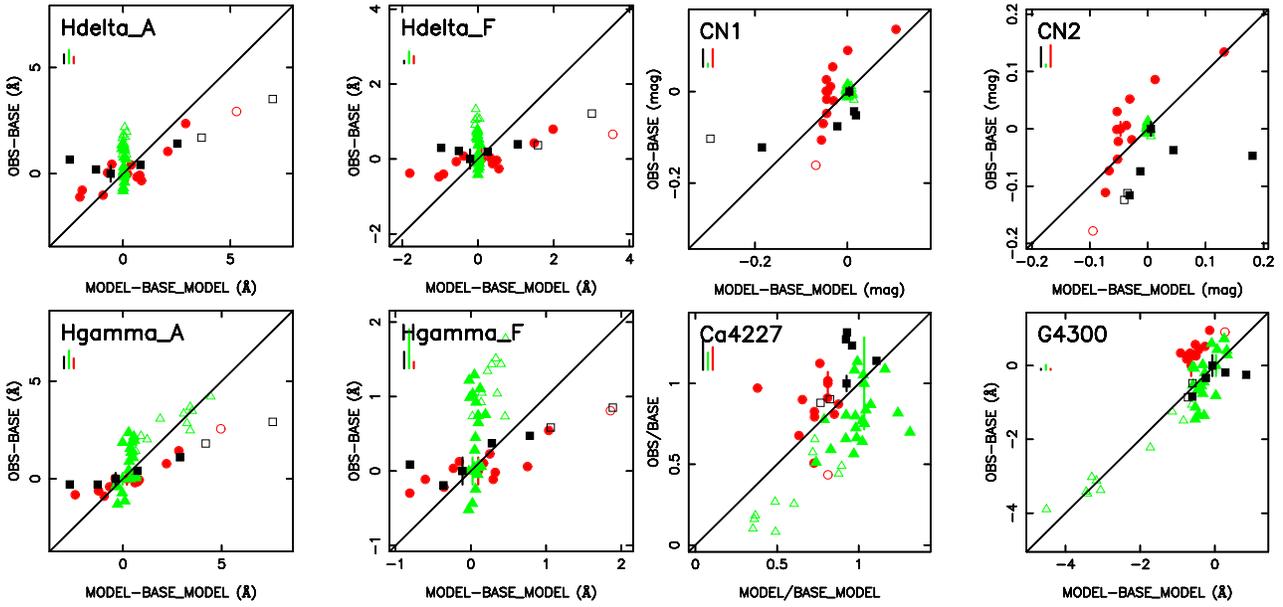

 \includegraphics[width=84mm, angle=0]{pgpage1Hlines_TRENDS_CGplusCnN.eps}
 \includegraphics[width=84mm, angle=0]{pgpage1CNOSensitive_TRENDS_CGplusCnN.eps}
 \caption{Testing the response functions of K05 by applying mean trends of
element abundance ratios. The assumed 
element abundance ratios are modified by relations seen for stars in the 
local disk of the Milky Way. Comparison of normalised empirical versus 
normalised theoretical line strengths for standard Lick indices sensitive 
to four H Balmer lines and to four CN sensitive indices, in the stellar 
photospheres. Symbols as in Fig.~1(a), with cool dwarfs (CD, black squares), 
turn-off stars (TO, green triangles) and cool giants (CG, red circles).}
\end{figure*}

\begin{table*}
\begin{center}
\begin{minipage}{60mm} 
\caption{Balmer line spectral index responses (changes in $\AA$) 
to overall metallicity changes (by a factor of 2). From tables 12 and 14 
in K05 and corresponding tables in H02.}
\begin{tabular}{llcc}
\hline
 & & K05 [Z/H] & H02 [Z/H] \\
\hline
 CD:     & H$\delta_A$ & -1.089    & -0.704    \\
         & H$\delta_F$ & -0.546    & -0.280    \\
         & H$\gamma_A$ & -1.381    & -0.432    \\
         & H$\gamma_F$ & -0.269    & -0.099    \\
 & & & \\
 CG:     & H$\delta_A$ & -1.533    & -2.112    \\
         & H$\delta_F$ & -1.252    & -0.640    \\
         & H$\gamma_A$ & -1.820    & -0.720    \\
         & H$\gamma_F$ & -0.534    & -0.132    \\
\hline
\end{tabular}
\end{minipage}
\end{center} 
\label{appendix_ZHBalmers}
\end{table*}

\bsp

\label{lastpage}

\end{document}